\documentclass[12pt]{article}

\usepackage{amsmath}
\usepackage{amssymb}
\usepackage{amsthm}
\usepackage{color}
\usepackage{appendix}
\usepackage{booktabs}
\usepackage[justification=centerfirst]{caption}
\usepackage{hyperref}
\usepackage{hypcap}
\usepackage{graphicx}
\usepackage{revsymb}
\usepackage{accents}

\begin{document}

%\title[Absorbing Media and the Schwarzschild Metric]{Ray Tracing Through Absorbing Dielectric Media in the Schwarzschild Spacetime}
\title{Ray Tracing Through Absorbing Dielectric Media in the Schwarzschild Spacetime}
\author{Adam Rogers \\ Department of Physics and Astronomy \\ University of Manitoba \\ Winnipeg, Manitoba, Canada \\ R3T 2N2 }
%\address{Department of Physics and Astronomy, University of Manitoba, Winnipeg, Manitoba, R3T 2N2, Canada}
%\ead{rogersaga@protonmail.com}

% https://orcid.org/0000-0003-2953-2054

\maketitle

\begin{abstract}
General Relativity describes the trajectories of light-rays through curved spacetime near a massive object. In addition to gravitational lensing, we include an absorbing dielectric medium given by a complex refractive index known as the Drude model. When absorption is included the eikonal becomes complex, with the imaginary part related to the absorption along a ray between emission and observation points. We extend results from the literature to include dispersion in the index of refraction. The complex Hamiltonian splits into a real part that describes the equations of motion and a constraint equation that governs the momentum loss in the system. We work in coordinates which are fully real, with a real metric in physical spacetime. We assume the dust and plasma distributions of the Drude matter to coincide and vary as a power-law $1/r^h$. We find that transmission requires $h>1$, otherwise exponential absorption occurs along ray paths. We use ray-tracing through strongly absorbing matter near the surface of the compact star, as well as specializing to a point-lens in the weak-field limit with weakly absorbing matter to generate potentially observable light curves for distant observers. In the appropriate limits, our theory reproduces results from the literature.
\end{abstract}

%\noindent{\it Keywords\/}: \textcolor{red}{gravitation - plasmas, dust, extinction, black hole physics, stars: neutron, pulsars: general}

% \submitto{Journal of Classical and Quantum Gravity}

\section{Introduction}
\label{sec:intro}

Ray tracing is a powerful procedure for mapping the behaviour of light rays around compact objects such as neutron stars and black holes that occupy complicated environments \cite{olegReview, olegShadow, erMao}. These objects are often surrounded by complicated magnetospheres \cite{magnetosphere1}, accretion flows \cite{fallback}, distributions of dust \cite{pageNS} and plasma \cite{shapiro}. Describing the path of rays through these regions requires an understanding of optics in both curved spacetime and dielectric materials \cite{synge1, OG_GR_plasma, volkerBook}. On this front, there has been a substantial amount of investigation into relativistic ray tracing around massive objects such as black holes and neutron stars immersed within a cold plasma medium \cite{oleg2, olegShadow, olegPlasmaTimeDelay, bozzaOleg}. There have also been studies on the optical effects of plasma on hypothetical exotic objects and within modified theories of gravity \cite{exotic1, exotic2, exotic3}. Generally objects lensed by both gravity and plasma show a variety of unique light curves \cite{olegHillsHoles}. However, a factor not considered in these literature studies is absorption. Generally, dust is relevant to observations of compact objects over all frequency ranges \cite{dustFreqs} due to intervening material along the line of sight to bright sources \cite{dustEnvironments, J119, magnetarDust1}.

In this work, we perform ray tracing through a spherically symmetric spacetime including a complex index of refraction \cite{drudeOrig, eikonal2}, which allows us to describe both refraction and absorption. This causes both the eikonal and the wave-vector (momentum) to become complex \cite{rayTracingAxions}. The imaginary part of the eikonal describes the absorption that a ray experiences along its trajectory. 

 The complex refractive index describes a ``Drude medium'', an absorbing dusty plasma, that envelops a non-rotating, compact object. The model has two limits. In one limit, the Drude medium reproduces the cold plasma index of refraction and contains previous results from the literature \cite{OG1, OG2, OG3}. In the other limit, the Drude medium behaves like a dielectric dust, which reduces the apparent intensity of the background flux. In this limit, the Drude model has been widely studied in terms of its description of X-ray scattering from a dusty halo \cite{drudeXRay, xrayDrude2}. Physically, dust grains effectively behave like small clouds of gas to X-rays, and thus low-angle scattering is described by a dielectric function. This work therefore extends the applicability of methods based on Synge's medium equation, which have previously focused on the low-frequency radio regime due to the effects of the cold plasma, which are unobservable at high frequencies far in excess of the plasma frequency \cite{synge1, volkerBook, CP1, CP2}. Our work also derives the weak-deflection limit using a power-law density for the dusty plasma halo. 

A complex index of refraction has been used in a cosmological setting with absorption in the optical metric as an alternative to the standard model cosmology \cite{reissPerlmutter, chenKantowski1, chenThesis}. Early studies were successful fitting the model to supernova redshift data without the need for a dark energy component \cite{chenKantowski2, dustCosmology}. However, by including Baryon Acoustic Oscillation data and carefully accounting for observational uncertainty in the supernova data set, it was later shown \cite{chen13} that cosmic opacity alone cannot independently account for cosmological observations and would still require both dark matter and dark energy components. However, in light of the Hubble tension in modern cosmology, interest remains in constraining the potentially observable effects of cosmic opacity \cite{ck20}. 

Regardless of opacity as a viable cosmological model, the work of Chen \& Kantowski \cite{chenKantowski1, chenThesis, chenKantowski2} also feature a study of the Schwarzschild spacetime imbued with a spatially-dependent absorbing optical medium. Our work will extend the results of Chen \& Kantowski to include dispersive media and demonstrate ray-tracing through both strongly and weakly absorbing media for gravitational lensing. 

We provide some background on Hamiltonian ray theory and Synge's medium equation in Section \ref{sec:theory}. Then we discuss opacity in the Schwarzschild background in section \ref{sec:opacity}. We will examine the general results and the spherically symmetric example explored by Chen \& Kantowski. Using our results we will discuss the Drude model, which describes a dispersive, spatially varying index of refraction, in Section \ref{sec:dustyPlasma}. We apply the Drude model and trace ray trajectories through strongly absorbing media in Section \ref{sec:rayTracingStrong} and explore the opposite case, the weak-deflection limit with weakly absorbing media, in Section \ref{sec:ML}. Finally, we discuss the linear Kramers-Kronig relationship in Section \ref{sec:KK} which allows the construction of physically coupled real and imaginary parts of the index of refraction. We discuss our results and open questions for future work in Section \ref{sec:disc} and, finally, summarize our findings in Section \ref{sec:conclusions}.

\section{Theory}
\label{sec:theory}

We follow the conventions used in the work of Bisnovatyi-Kogan and Tsupko \cite{olegRadioLens}. We define the metric signature $(-,+,+,+)$. We use the Einstein summation convention with Latin indices $a$, $b$, $c$, etc. for summing over $4$-vector quantities and Greek indices $\alpha$, $\beta$, $\gamma$, etc. for summing over the $3$-vector spatial components only. We adopt the coordinates $x^a=(t$, $r$, $\theta$, $\phi$), and index using the coordinate names. For example, the equatorial angle $\phi$ is specified by the coordinate $x^\phi$ and the corresponding angular momentum is $p_\phi$. Square brackets represent antisymmetrization of the indices they contain. We will use caligraphic fonts $\mathcal{R}$ and $\mathcal{I}$ as labels to denote the real and imaginary parts of quantities. We will put these labels in the opposite position from tensor indices, such as $a_{ij}^\mathcal{R}$ for the real part of the $ij$ component of $a$. From here on we define an overdot as shorthand for the ordinary derivative with respect to the affine parameter. An overbar on a parameter means that it has been acted upon by the complex optical metric. A star $\accentset{\ast}{\xi}$ stands for complex conjugation. We are rigorous in retaining all constant factors in our analysis, but will use scaled units with $c=G=\hbar=1$ for our numerical calculations. 

The starting point of the analysis carried out by Chen \& Kantowski \cite{chenKantowski1} begins with Maxwell's equations, far from any sources
\begin{equation}
\nabla_b F^{a b} = 0
\label{eq:max1}
\end{equation}
and
\begin{equation}
\nabla_{[a}F_{b c]} = 0,
\label{eq:max2}
\end{equation}
where $F_{a b}$ is the antisymmetric Faraday field tensor. The Maxwell equations \ref{eq:max1} and \ref{eq:max2} have solutions that describe monochromatic waves. The four-potential $\mathcal{A}_\alpha$ is used to write the electromagnetic field tensor,
\begin{equation}
    F_{a b} = 2\nabla_{[a} \mathcal{A}_{b ]}.
\end{equation}
Additionally, the potential also obeys the source-free wave equation
\begin{equation}
\nabla_a \nabla^a \mathcal{A}^b -R_a^b \mathcal{A}^a=0    
\label{eq:maxWave}
\end{equation}
where we are justified in ignoring the terms containing the Ricci tensor due to the large scale of spacetime curvature compared to a single wavelength. The Maxwell equations and wave equation are supplemented by the Lorenz gauge condition, 
\begin{equation}
    \nabla_a \mathcal{A}^a = 0.
    \label{eq:max3}
\end{equation}
The geometric optics limit describes a travelling wave solution with a rapidly changing real phase and a slowly changing complex amplitude. We define the wavevector as the gradient of the phase,
\begin{equation}
    k_a = \partial_a S.
    \label{eq:kEikonal}
\end{equation}
Next we introduce the WKB approximation with geometric expansion parameter $\lambdabar$, such that
\begin{equation}
\mathcal{A}^{a} =  \left[ \xi^a + \lambdabar \eta^a +O(\lambdabar^2) \right] e^{i\frac{S(x^a)}{\lambdabar}} 
\end{equation}
the phase of the wave is given by a real scalar function $S(x^a)$, with $\xi^a$ and $\eta^a$ complex $0^\text{th}$ and $1^\text{st}$ order amplitudes. The expansion parameter  $\lambdabar$ is identified with the reduced wavelength \cite{reducedWavelength}. In flat space the reduced wavelength is
\begin{equation}
\lambdabar = \frac{\lambda}{2 \pi} = \frac{c}{\omega(x^a)}
\label{eq:reducedWavelength}
\end{equation} 
however, the exact details of the coordinate dependence are found using the gravitational redshift, which is represented by $\omega(x^a)$ on the right hand side. The frequency $\omega(x^a)$ is measured in the instantaneous rest frame of the medium \cite{oleg24}. As described in \cite{mtw}, the reduced wavelength defines the classical distance of closest approach for a photon with one unit of angular momentum.

The geometric optics dispersion relation is found using the WKB ansatz in the wave equation and the Lorenz gauge condition, eqs. \ref{eq:maxWave} and \ref{eq:max3}, and collecting terms of each order in the expansion parameter $\lambdabar$, which must each vanish individually. The wave equation provides the eikonal equation for light rays in the vacuum,
\begin{equation}
 k^a k_a = 0,
\end{equation} 
and gives the transport equation for the amplitude vector, 
\begin{equation}
    k^b\nabla_b \xi_a +\frac{1}{2}\xi_a\nabla_b k^b = 0.
    \label{eq:propEqn}
\end{equation}
The Lorenz gauge condition shows the amplitude vector is normal to the propagation direction 
\begin{equation}
    \xi^a k_a = 0.
\end{equation}
The divergence parameter of null rays is given as 
\begin{equation}
    \Theta = \frac{1}{2} \nabla_a k^a,
\end{equation}
and we define a tick over a variable as a differential operator, in the spirit of notation used in \cite{chenKantowski1},
\begin{equation}
    \xi_a' = k^b \nabla_b \xi_a.
\end{equation}
Putting these pieces together, we find the vector amplitude transport equation in the vacuum, 
\begin{equation}
    \left(\xi^a \xi_a \right)' + 2 \left(\xi^a \xi_a \right) \Theta  = 0.
    \label{eq:vacAmpXPort}
\end{equation}

The optical metric describes the effects of the spatially varying index of refraction in terms of an effective vacuum metric \cite{synge1, gordon1923, galloOpticalMetric}. Formally, Gordon's optical metric is 
\begin{equation}
    \bar{g}_{ab} = g_{ab} + \frac{1}{c^2} \left( 1-\frac{1}{N^2}\right)u_a u_b
\end{equation}
with $g_{ab}$ the physical spacetime metric and $\bar{g}_{ab}$ the optical metric. Just as the metric is invariant with respect to conformal transformations, the modified Maxwell equations in Gordon's optical spacetime are \cite{chenKantowski2}
\begin{equation}
    \partial_{[a}\bar{F}_{bc]}=0
\end{equation}
and
\begin{equation}
    \bar{\nabla}_b\left( e^{2\Phi_Z}\bar{F}^{ba} \right)=0
\end{equation}
with the conformal factor given by the reciprocal of the optical impedance,
\begin{equation}
    e^{2\Phi_Z} = \sqrt{\frac{\varepsilon}{\mu}}. 
\end{equation}

Suppose the optical metric is static, stationary and spherically symmetric. The line element for the optical spacetime is, 
\begin{equation}
    d\bar{s}^2 = -\frac{A(r)}{N(r)^2}(cdt)^2 +  B(r)dr^2 + r^2(d\theta^2 + sin^2\theta d\phi^2).
\end{equation}
In addition to the central mass, we assume spacetime is filled with a spherically symmetric distribution of absorbing optical fluid that is described by the complex refractive index
\begin{equation}
    N(r)=n(r) + i \kappa(r).
    \label{eq:Ncomplex}
\end{equation}
We assume the optical properties of the fluid are a function of radial distance only. The optical material is static and fills all of space. It is at rest with respect to the mass $M$. We write the $4$-velocity of the material as  
\begin{equation}
    u^a=(c\sqrt{-g^{tt}},\vec{0}),
    \label{eq:mediumVelocity}
\end{equation}
The assumption of a static distribution of fluid is commonly used in the lensing literature (see for example \cite{olegReview} for a thorough review), however recent analytical work has included dynamic media \cite{oleg24}. For now, let us neglect dispersion effects in the index of refraction such that $N(r)$ does not depend on frequency \cite{chenKantowski1, chenKantowski2, galloOpticalMetric}, however we will return to this point in a later section. 

Following the same reasoning as the vacuum case earlier, \cite{chenKantowski1} establishes the vector transport equation eq.\ref{eq:vacAmpXPort} now acquires an extra term that depends on the derivative of the optical reciprocal impedance, 
\begin{equation}
    \left(\bar{\xi}^a\xi_a\right)' + 2 \left(\bar{\xi}^a\xi_a\right) \left( \Theta  + \Phi'_Z \right) = 0.
    \label{eq:propODE1}
\end{equation}
The expansion parameter for the propagation direction in the optical metric is 
\begin{equation}
    \Theta = \frac{1}{2}\bar{\nabla}_a \bar{k}^a.
\end{equation}
The geodesics of the optical metric are given by the null condition 
\begin{equation}
    \bar{k}^a k_a =0
    \label{eq:nullCond1}
\end{equation}
with the overbar on $k$ denoting it has been acted on by the optical metric. Eq. \ref{eq:nullCond1} is the geometric optics dispersion relation, the Hamilton-Jacobi equation for the phase $S$ \cite{spinHall}. Using Eq. \ref{eq:nullCond1} along with Eq. \ref{eq:kEikonal}, we recover the geodesic equation in the optical spacetime
\begin{equation}
    \bar{k}^b \bar{\nabla}_b \bar{k}^a = 0.
\end{equation}
With the Lorenz gauge, we have 
\begin{equation}
    \bar{k}^a \xi_{a} = k_a \accentset{\ast}{\bar{\xi}}^{a} = 0.
    \label{eq:ka2}
\end{equation}
 This is the homogeneity condition, which expresses the orthogonality of the wave vector $k_a$ and amplitude vector $\xi_{a}$. The amplitude vector is spacelike and transverse to the waves propagation direction as seen by the optical fluid, which we call $\hat{k}^a$ such that $\xi_{a}\hat{k}^a = 0$. This further implies that $\xi_{a} u^a = 0$. Physically, the amplitude vector is related to the real scalar intensity $\xi=(\xi_a \accentset{\ast}{\bar{\xi}}^a)^\frac{1}{2}$ as
\begin{equation}
    \xi_{a} = \xi f_a
\end{equation}
where $f_a$ is a unit-complex polarization vector \cite{spinHall}. The WKB expansion gives the transport equation for the complex amplitude vector, 
\begin{equation}
    \bar{k}^b \bar{\nabla}_b \xi_{a} + \frac{1}{2}\xi_{a} \bar{\nabla}_b \bar{k}^b = 0.
\end{equation}
Additional transport equations can be derived for the scalar intensity $\xi$ and the polarization vector $f_b$:
\begin{equation}
    \bar{\nabla}_a (\xi \bar{k}^a) = 0,
\end{equation}
\begin{equation}
    \bar{k}^b \bar{\nabla}_b f_a = 0.
\end{equation}

Waves that obey the condition given in Eq. \ref{eq:ka2} are called homogeneous waves \cite{eikonal2}, which are defined by having an amplitude vector that coincides with a surface of constant phase. The  covariant wave vector for a homogeneous wave is, 
\begin{equation}
    k_a = -(S_{,b}u^b) \left[u_a + N\hat{k}_a \right]    \label{eq:homogenousWave1}
\end{equation}
and the complex contravariant wave vector acted upon by the optical metric,
\begin{equation}
    \bar{k}^a = -(S_{,b}u^b) N \left[ Nu^a + \hat{\bar{k}}^a \right].
\label{eq:homogenousWave2}
\end{equation}
The unit space-like vector $\hat{k}_a$ with $\hat{\bar{k}}^a \hat{k}_a > 0$, is the wave propagation direction that is measured by an observer moving with the fluid. Since it contains the index of refraction, the wave-vector (momentum) is naturally complex. Due to the connection between momentum and the eikonal (eq. \ref{eq:kEikonal}), this leads to the eikonal inheriting a complex nature as well. Let us write the complex eikonal as
\begin{equation}
    S(x^a) = S_\mathcal{R}(x^\alpha) + i S_\mathcal{I}(x^a).
\end{equation}
An observer travelling along with the fluid $u^a$ relates the eikonal with the local wave period $T$ of a transmitted wave, 
\begin{equation}
    -(S_{,b} u^b) = \lambdabar\left( \frac{2 \pi}{cT}\right) 
    \label{eq:ckPeriod1}
\end{equation}
\begin{equation}
    \frac{2\pi}{\lambda} + i \frac{\alpha_D}{2}  = N\left( \frac{2 \pi}{cT} \right).
    \label{eq:ckPeriod2}
\end{equation}
Finally, Chen \& Kantowski define the time-averaged 4-flux $S_F$ using the Poynting vector. With a complex index of refraction, the flux contains an extra enveloping function that describes energy loss due to absorption \cite{chenKantowski1},
\begin{equation}
    S_F^a = e^{-2 \frac{S_\mathcal{I}}{\lambdabar}}\left[ \frac{c}{8 \pi}(\xi_a \accentset{\ast}{\bar{\xi}}^a)|S_{,b}u^b|^2 Re\left\{ \sqrt{\frac{\epsilon}{\mu}}\right\}\left(Re\{ N \}u^a + \hat{k}^a \right) \right].
\end{equation}
The exponential factor depends only on the imaginary part of the eikonal. This factor represents the flux transferred from the wave to the medium \cite{chenKantowski1, chenKantowski2}. In terms of the complex eikonal, the absorption factor is given by eq. \ref{eq:ckPeriod2}, 
\begin{equation}
\alpha_D(x^a) = 2 \frac{S_\mathcal{I}(x^a)}{\lambdabar} 
 = 2 \kappa(r) \frac{\omega(r)}{c}
\end{equation}
where the frequency $\omega(r)$ is affected by the gravitational redshift (eq.\ref{eq:reducedWavelength}). The optical depth is defined as the integral along the ray path \cite{carroll}
\begin{equation}
    \tau(x^a) = \int_{\ell_0}^\ell \alpha_D(r)d\ell, 
\end{equation}
and the specific intensity of radiation between the reception point $\ell_0$ and the emission point $\ell$ in the emitters frame is \cite{rybickiLightman04, vincent11},
\begin{equation}
    I =  I_0 \exp\left( - \tau \right).
\end{equation}
where $I$ is the specific intensity in the observer's frame and $I_0$ is the constant intensity emitted by the source. The integral is evaluated along the path $d\ell$, the line of sight from the observer to the source.

\section{Opacity in the Schwarzschild Spacetime}
\label{sec:opacity}

Radially outgoing rays in the optical Schwarzschild spacetime were studied by Chen \& Kantowski \cite{chenKantowski1}. We will review their results as they relate to the absorption coefficient to motivate our study of the equations of motion. 

Let us describe a compact object surrounded by absorbing matter using the Schwarzschild metric. We also assume a spherically symmetric complex index of refraction as in \ref{eq:Ncomplex}. As seen from the null condition (eq. \ref{eq:nullCond1}), light rays in the optical metric travel along null geodesics. In contrast, light rays travel along time-like geodesics in the physical spacetime, which is also spherically symmetric, static and stationary, with the line element
\begin{equation}
    ds^2 = -A(r)(cdt)^2 +  B(r)dr^2 + r^2(d\theta^2 + \sin^2\theta d\phi^2).
\end{equation} 
In terms of the Newtonian gravitational potential $V_G(r)$, the Schwarzschild metric has 
\begin{equation}
g_{tt}(r) = A(r)=e^{2\Phi(r)} = 1-\frac{2 }{c^2}V_G(r) = 1-\frac{2GM}{c^2r}    
\end{equation}
and
\begin{equation}
g_{rr}(r) = B(r)=e^{2\Psi(r)} = \frac{1}{ 1-\frac{2 }{c^2}V_G(r) } = \frac{1}{ 1-\frac{2 G M}{c^2 r}}
\end{equation}
with the compact object mass $M$ at the coordinate origin. We use $A(r)$ and $B(r)$ for the lensing calculations \cite{olegGeneral} in Section \ref{subsec:disp}, whereas for now we use the exponential forms to expedite comparison with Chen \& Kantowski. 

Let the time element of the wave vector be called $k_t$, which is constant due to the static nature of the metric. The value of this constant is fixed from the wave-vector at spatial infinity $r \rightarrow \infty$, where the curvature vanishes and spacetime is flat (Minkowski). This gives us the constant
\begin{equation}
    k_t=-\frac{\omega_\infty}{c}
    \label{k0constant}
\end{equation}
with $\omega_\infty$ the frequency at spatial infinity. With this identification we establish the relationship between wave frequency and the velocity of an observer at rest with respect to the dielectric medium. In combination with the medium velocity, the wave vector gives us the gravitational redshift, 
\begin{equation}
    u^a k_a = c \sqrt{ -g^{tt} }k_t = -\sqrt{ -g^{tt} }\omega_\infty = -\omega(r). 
    \label{eq:vk}
\end{equation}
This is the frequency that is measured by an observer comoving with the optical fluid. Finally, consider an outwardly-directed radial trajectory \cite{chenKantowski1, chenKantowski2},
\begin{equation}
k_a = \left(-1, N e^{-\Phi + \Psi}, 0, 0 \right).
\label{eq:kCK1}
\end{equation}
The photon energy can always be normalized to $k_t=-1$ by adjusting the affine parameter by a constant factor \cite{dempsey}. For such a radially directed ray, the complex eikonal is 
\begin{equation}
   S = -ct + \int_0^r n(r')e^{-\Phi+\Psi}dr' + i\int_0^r \kappa(r') e^{-\Phi+\Psi}dr'. 
   \label{eq:fullComplexEikonal1}
\end{equation}
The equations related to the observed period (eqs \ref{eq:ckPeriod1} and \ref{eq:ckPeriod2}) are also modified in the presence of an absorbing medium. We assume a constant and steady source emitting radiation. The above expressions simplify to give the relationships between the frequency, wavelength, attenuation factor and geometric optics expansion parameter: 
\begin{equation}
    \omega(r)=2\pi\nu = \frac{2 \pi}{T} =e^{-\Phi} \frac{c}{\lambdabar}= \omega_0 e^{-\Phi} 
\end{equation}
which is the usual gravitational redshift in Schwarzschild spacetime. This expression shows that the frequency evolves due to the radial motion in the curved spacetime. The real part of eq \ref{eq:ckPeriod2} demonstrates the effect of the real part of the index of refraction on the wavelength, 
\begin{equation}
    \lambda(r) = 2 \pi \lambdabar \frac{e^\Phi}{n(r)}
\end{equation}
and the imaginary part of eq \ref{eq:ckPeriod2} relates to the attenuation coefficient, 
\begin{equation}
    \alpha_D(r) = \frac{2 \kappa(r) e^{-\Phi}}{\lambdabar}
\end{equation}
and the optical depth 
\begin{equation}
    \tau(r) = \frac{2}{\lambdabar}\int_0^r \kappa (r')e^{-\Phi+\Psi} dr'. 
\end{equation}
Using eq. \ref{eq:fullComplexEikonal1}, we reproduce the result from \cite{chenKantowski1} for the optical depth, 
\begin{equation}
    \tau(r) = \int_0^r \alpha_D(r') e^\Psi dr'.
\end{equation}

With the results of Chen \& Kantowski established using the optical spacetime \cite{chenKantowski1}, we will now turn to the equations of motion determined by the optical Hamiltonian using Synge's medium equation \cite{synge1}.

\subsection{Dispersive Media in Physical Spacetime}
\label{subsec:disp}

The optical metric describes all refractive effects in terms of an ``effective spacetime'' whose curvature accounts for the effect of an optical medium. It is well-established that the equations of motion of both the vacuum optical metric and the physical metric with absorbing optical media correspond to one another. 

The optical metric substantially simplifies the equations of motion, however the approach has a significant drawback in that it cannot accommodate dispersive effects \cite{synge1, galloOpticalMetric}. When dispersive media are used, the optical metric becomes frequency dependent. Hence, it is no longer a true metric since each frequency ray would require its own unique metric description. However, as pointed out in \cite{galloOpticalMetric}, there are optical metric alternatives to the Gordon metric that can incorporate dispersive media. Moreover, Synge's medium equation in physical spacetime has no such shortcoming itself for including dispersive media.

Our strategy going forward is to include a dispersive, absorbing optical medium within Schwarzschild spacetime. Solving the equations of motion with the medium equation allows us to describe novel lensing phenomena which include potentially observable effects. The analysis of Chen \& Kantowski \cite{chenKantowski1, chenKantowski2} establishes the quantitative relationship between the imaginary part of the eikonal and the absorption coefficient. Using this result with the medium equation allows us to describe absorption with an index of refraction that contains frequency dependence.

It will be advantageous for us to define the index of refraction in terms of the real and imaginary parts of the dielectric susceptibility $\chi_\mathcal{R}(\omega, r)$ and $\chi_\mathcal{I}(\omega, r)$, respectively. We then have 
\begin{equation}
  N^2(\omega, r) = \frac{\varepsilon(\omega, r)}{\varepsilon_0} = 1 + \chi_\mathcal{R}(\omega, r) + i \chi_\mathcal{I}(\omega, r).
\label{eq:complexIndRef}
\end{equation}
The ratio is called the \textit{relative electric permittivity}, with the permittivity $\varepsilon(\omega, r)$ and the permittivity of the vacuum $\varepsilon_0 \approx 1.85 \times 10^{-12}$ F/m \cite{eikonal2}. Physically, the induced electric dipole moment per unit volume $\vec{P}$ inside of a dielectric is given by the electric field $\vec{E}$ with susceptibility as linear coefficient, $\vec{P}=\varepsilon_0\chi \vec{E}$. Equating the complex refractive index to the susceptibility directly, we find
\begin{equation}
    \chi_\mathcal{R}(\omega, r) = n^2 - \kappa^2 - 1
\end{equation}
and
\begin{equation}
    \chi_\mathcal{I}(\omega, r) = 2n\kappa.
    \label{eq:opticsImag}
\end{equation}
Dealing with the complex components of the square of the index of refraction in terms of the susceptibilities will simplify Synge's approach to ray tracing in GR. The real and imaginary susceptibilities are dependent on one another and cannot simply be chosen arbitrarily. The two parts of the susceptibility are linked through the Kramers-Kronig relations that enforce causality in complex systems. Given some arbitrary function for the real susceptibility, the Kramers-Kronig relations can be linearized \cite{linKK1} and used as a method for generating the corresponding physically relevant imaginary part (or a given imaginary part can be used to generate a corresponding real partner) \cite{linKK2, linKK3}. We will discuss the linear Kramers-Kronig relations in Section \ref{sec:KK}.

The trajectories of massless particles are described in terms of an affine parameter $\sigma$. The affine parameter is defined so that the coordinate derivative along the path with respect to the affine parameter gives the $4$-momentum,
\begin{equation}
    \frac{dx^a}{d\sigma}=p^a
\end{equation}
The affine parameter is normalized to measure proper distances in the local rest frame of the observer. From the dispersion relation (eq. \ref{eq:nullCond1}) we define the Hamiltonian,
\begin{equation}
    H(x^a, p_a) = \frac{1}{2} g^{ab} p_a p_b = 0
    \label{eq:hamFull1}
\end{equation}
and Hamilton's equations of motion,
\begin{equation}
    \dot{x}^a = \frac{dx^a}{d\sigma}= \frac{\partial H}{\partial p_a}
    \label{eq:hamEOM1}
\end{equation}
\begin{equation}
    \dot{p}_a = \frac{dp^a}{d\sigma}= - \frac{\partial H}{\partial x^a}.
    \label{eq:hamEOM2}
\end{equation}
Using the Hamiltonian (eq. \ref{eq:hamFull1}) above, we find 
\begin{equation}
    \dot{x}^a = Re\left\{ g^{a b}p_a \right\}
    \label{eq:hamEOM1old}
\end{equation}
\begin{equation}
    \dot{p}_a = -\frac{1}{2} \partial_a g^{b c}p_b p_c = \Gamma^e_{b a} p_e p^b.
    \label{eq:hamEOM2old}
\end{equation}
These relationships can be manipulated to reproduce the geodesic equation, demonstrating once again that light rays in vacuum travel along null geodesics \cite{mtw}. 

The medium equation \cite{synge1} provides the Hamiltonian including curved spacetime and a dispersive lossy medium:  
\begin{equation}
H(x^a, p_a, u^a)=\frac{1}{2} \left[ g^{ab} p_a p_b - \frac{(N^2-1)}{c^2}(u^a p_a)^2\right] = 0.
\label{eqGH1}
\end{equation}
With the complex index of refraction given in eq. \ref{eq:Ncomplex}, this expression becomes
\begin{equation}
H(x^a, p_a, u^a)=\frac{1}{2} \left[ g^{ab} p_a p_b - \frac{\chi_\mathcal{R}}{c^2}(u^a p_a)^2 - i\frac{\chi_\mathcal{I}}{c^2}(u^a p_a )^2 \right] = 0.
\label{eqPermH1}
\end{equation}
The extra imaginary term in the Hamiltonian depending on $\chi_\mathcal{I}$ is related to the medium absorption. In the limit $\chi_\mathcal{I} \rightarrow 0$ this term vanishes and the expression reduces to the lossless case studied in \cite{olegGeneral} for the ray trajectories of radiation through a dispersive medium in a curved background spacetime. In this case, the rays follow trajectories similar to the paths they would follow in the case of a non-absorbing medium, while the imaginary susceptibility affects the amplitude change due to the absorption of the medium. Note that naively it looks impossible for the Hamiltonian to vanish over the ray trajectory for the imaginary part as it is shown in equation \ref{eqPermH1} at first glance, since only a single imaginary term appears there. However, the situation is substantially more complicated since the radial momentum $p_r$ must also have an imaginary part that contributes to the complex eikonal. These contributions will allow both real and imaginary parts of the Hamiltonian in eq. \ref{eqPermH1} to vanish independently. We note that the coordinates in the problem are physically real. The complex nature of the momentum using the Hamiltonian is due to the link between momentum and the eikonal. Since the eikonal is now complex, this is also reflected in the momentum. 

Using the index of refraction in terms of the dielectric susceptibility, and the medium velocity (eq. \ref{eq:vk}), the Hamiltonian becomes
\begin{equation}
H(x^a, p_a)=\frac{1}{2} \left(g^{\alpha \beta}p_\alpha p_\beta + g^{tt} p_t^2 + g^{tt} p_t^2 \chi_\mathcal{R} \right) + i  \frac{1}{2} \chi_\mathcal{I} g^{tt} p_t^2.
\label{ham1}
\end{equation} 
we evaluate the equations of motion using eqs. \ref{eq:hamEOM1} and \ref{eq:hamEOM2}. The time component is
\begin{equation}
 \dot{x}^t = c\dot{t} = Re \left\{g^{tt}p_t \left(1+\chi_\mathcal{R}+i \chi_\mathcal{I} \right) \right\}.  
 \label{eq:TimeEq}
\end{equation}
and for the spatial components we find \begin{equation}
    \dot{x}^a = Re \left\{ g^{a b}p_b \right\} .
    \label{eq:xAlphaDef}
\end{equation}
This expression gives
\begin{equation}
 \dot{r}=\frac{Re\left\{p_r\right\}}{B}. 
\label{dr1}
\end{equation}
The time component of the momentum is the initial energy, which is real and constant $p^t=E_\infty/c$,  
\begin{equation}
    \dot{p}_t=0
    \label{eq:tAlphaDef}
\end{equation}
due to the Hamiltonian being independent of time. Similarly, since the Hamiltonian is independent of $\phi$, we also have the angular momentum $p_\phi=L$ constant, and conserved over the trajectory $\dot{\phi}=0$, $\dot{p}_{\phi}=0$. Moreover, due to spherical symmetry any orbital plane through the mass $M$ is equivalent. Without loss of generality we follow trajectories in the equatorial plane and set $\theta=0$, $\dot{\theta}=0$, $p_\theta=0$ and $\dot{p}_\theta=0$. 

The radial momentum is most easily found by the vanishing of the Hamiltonian and the constant photon energy, 
\begin{equation}
 p_r = \pm \left[ \frac{B}{A} p_t^2 \left( 1 + \chi_\mathcal{R} \right) -  \frac{B}{r^2} p_\phi^2 + i \frac{B}{A} p_t^2 \chi_\mathcal{I}  \right]^\frac{1}{2}.
\label{radMom}
\end{equation}
Let us return for a moment to the radially outgoing rays that were used earlier (eq. \ref{eq:kCK1}). Radially directed photons have no angular momentum so $p_\phi=0$. Then, normalizing the time component to $p_t=-1$ \cite{dempsey} leads eq. \ref{radMom} to give the radial momentum that reproduces the wave vector used by Chen \& Kantowski \cite{chenKantowski1}. 

Once again we make use of the equations of motion (eqs. \ref{eq:hamEOM1} \& \ref{eq:hamEOM2}) for the change in radial momentum which gives 
\begin{equation}
 \dot{p}_r = \frac{1}{2} \left[ \frac{p_r^2}{B^2} \frac{dB}{dr} + 2\frac{p_\phi^2}{r^3} + p_t^2 \frac{d}{dr}\left(\frac{1+\chi_\mathcal{R}}{A} \right)\right] +i\frac{1}{2} p_t^2 \frac{d}{dr}\left(\frac{\chi_\mathcal{I}}{A} \right)  
\label{dotRadMom}
\end{equation}
The equations of motion impose strong constraints on the eikonal, which takes the form 
\begin{equation}
    S(x^a,k_a) = - E_\infty t + L\phi + S_r(r)
\end{equation}
with the polar angle term $S_\theta(\theta)=0$ due to symmetry. The radial part of the Eikonal is given by 
\begin{equation}
    S_r(r) = \int_{r_0}^r p_r dr.
    \label{eq:eikonalSR1}
\end{equation}
Using eqs. \ref{radMom} and \ref{eq:eikonalSR1} we find the radial eikonal, 
\begin{equation}
S_r(r)= \pm \int_{r_0}^r \left[ \frac{B}{A} p_t^2 \left( 1 + \chi_\mathcal{R} \right) -  \frac{B}{r^2} p_\phi^2 + i \frac{B}{A} p_t^2 \chi_\mathcal{I}  \right]^\frac{1}{2} dr
\label{eq:Seik}
\end{equation}
It is the imaginary part of this integral that controls the absorption profile. However, this expression is too complicated to solve analytically and must be evaluated numerically under general circumstances. This approach solves the equations of motion in terms of a complex radial eikonal. However, as can be done for all imaginary systems, we can re-write the expressions in terms of purely real variables. This will give us additional insight into the effects of lossy media on ray trajectories. 

Let us return once more to the complex Hamiltonian (eq. \ref{ham1}) and consider the effect of a complex wave vector. Suppose we substitute
\begin{equation}
    p_r = p_r^\mathcal{R} + i p_r^\mathcal{I}
    \label{eq:complexMom1}
\end{equation}
The Hamiltonian splits into real and imaginary parts
\begin{equation}
H_\mathcal{R}(x^a, p_a)=\frac{1}{2} \left(g^{\phi \phi}p_\phi^2 + g^{rr}(p_r^\mathcal{R})^2 - g^{rr}(p_r^\mathcal{I})^2 + g^{tt} p_t^2 + g^{tt} p_t^2 \chi_\mathcal{R} \right)=0
\label{eq:hamReal}
\end{equation} 
and
\begin{equation}
H_\mathcal{I}(x^a, p_a)=   g^{rr} p_r^\mathcal{R} p_r^\mathcal{I} + \frac{1}{2} \chi_\mathcal{I} g^{tt} p_t^2 = 0.
\label{eq:hamImag}
\end{equation} 
These equations give us a real Hamiltonian that is a modified version of the medium equation, which can be seen by writing 
\begin{equation}
    H_\mathcal{R} = H_{orig} + H_{corr}=0
\end{equation}
with $H_{orig}$ the original unmodified Hamiltonian, as in eq. \ref{eq:hamFull1}, and
\begin{equation}
    H_{corr}(x^a, p_a) = -\frac{1}{2}g^{rr}(p_r^\mathcal{I})^2.
\end{equation}
The real Hamiltonian $H_\mathcal{R}$ represents the usual vacuum Hamiltonian modified by an extra loss of momentum in the radial direction $H_{corr}$. The square of the real component of the ray momentum $(p_r^\mathcal{R})^2$ is partially cancelled off by the square of the imaginary portion $(p_r^\mathcal{I})^2$. Physically, $p_r^\mathcal{I}$ is the direction in which momentum loss occurs.

From the imaginary Hamiltonian eq. \ref{eq:hamImag}, we find the cross-term between real and imaginary parts of the radial momentum depends on the imaginary susceptibility 
\begin{equation}
g^{rr} p_r^\mathcal{R} p_r^\mathcal{I} =- \frac{1}{2} \chi_\mathcal{I} g^{tt} p_t^2.
\label{eq:hamImag2}
\end{equation}
The orthogonality condition from the imaginary part of the Hamiltonian eq. \ref{eq:hamImag2} can be satisfied in three possible ways. First, the condition is satisfied trivially if the wave vector is real ($k_a^\mathcal{I}=0$), which also implies a loss-free medium with vanishing imaginary susceptibility ($\chi_\mathcal{I}=0$). A more interesting second possibility is that the eikonal is complex but we instead restrict ourselves to a real index of refraction (again $\chi_\mathcal{I}=0$). When absorption is neglected, the real $p_r^\mathcal{R}$ and imaginary $p_r^\mathcal{I}$ parts of the momentum must be perpendicular to one another \cite{kravtsov99}
\begin{equation}
g^{rr} p_r^\mathcal{R} p_r^\mathcal{I} = 0    
\end{equation}
or, in terms of the imaginary part of the eikonal,
\begin{equation}
g^{rr} p_r^\mathcal{R} S^\mathcal{I}_{;r} = 0. 
\end{equation}
In the absence of absorption the ray trajectory is within a plane of constant amplitude $S_\mathcal{I}$, ie a plane within which the imaginary portion of the eikonal is constant. With no change in the wave amplitude along this trajectory, no loss occurs for this ray. The third and most general case for satisfying the constraint (eq. \ref{eq:hamImag2}) involves both a complex wave vector and complex index of refraction. When the imaginary susceptibility does not vanish, the condition eq. \ref{eq:hamImag2} can only be satisfied if the real and imaginary parts of the wave vector are oriented in some particular direction determined by the imaginary part of the medium susceptibility. Since the susceptibility varies throughout space $\chi_\mathcal{I}=\chi_\mathcal{I}(r)$ the orientation of the phase and amplitude planes vary with respect to one another. Since planes of constant amplitude no longer coincide with the wave-vector, transit of a photon along the trajectory causes a change in amplitude. The non-vanishing component of $p_r^\mathcal{I}$ along the trajectory cancels off a portion of the radial momentum $p_r^\mathcal{R}$. Thus, waves in lossy media show a combination of travelling and evanescent behaviour. The degree of evanescent behaviour is controlled directly by the imaginary susceptibility of the medium, $\chi_\mathcal{I}$.

The paths given by the real part of the Hamiltonian define trajectories that are perpendicular to the surfaces of constant phase $p_r^\mathcal{R}$. As the wave traverses the absorbing dielectric, the angle between $p_r^\mathcal{R}$ and $p_r^\mathcal{I}$ varies. This represents the orientation of the phase and amplitude planes dynamically evolving along the ray path determined by the spatial distribution of absorbing material. Using Hamiltons equations of motion gives the evolution equation for the momentum loss component, 
\begin{equation}
     \frac{d p_r^\mathcal{I}}{d \sigma} =  \dot{p}_r^\mathcal{I}  = - \frac{\partial H_\mathcal{I}}{\partial r},
\end{equation}
we write the rate of change
\begin{equation}
    \frac{d p^\mathcal{I}_a}{d \sigma} = -g^{r r}_{,a}p^\mathcal{R}_r p_r^\mathcal{I} + \frac{p_t^2}{2}\left( \chi_\mathcal{I}g^{tt}_{,a} + g^{tt}\chi^{\mathcal{I}}_{,a} \right) 
    \label{eq:dpI}
\end{equation}
for the $a=r$ component. This expression, along with the modified Hamiltonian, $H_\mathcal{R}$, provides a fully equivalent system that reproduces the solution to the complex ray tracing problem using only real quantities. Using either this set of equations, or the optical metric (eq. \ref{eq:hamFull1}) produces identical results. However, as previously stated, the medium equation has no restriction describing dispersive media with a frequency-dependent index of refraction. 

The rate of change of radial momentum component is given below with both real and imaginary parts separated and factored for clarity. Replacing the metric components, we have
\begin{equation}
\dot{p}_r^\mathcal{R} =  -\frac{p_r^2}{2} \frac{r_g}{r^2} +  \frac{p_\phi^2}{r^3} - \frac{\hbar^2}{2 c^2}\frac{ \omega_\infty^2}{\left(1-\frac{r_g}{r} \right)^2}\frac{r_g}{r^2}\left[ 1+\chi_\mathcal{R}- \left( 1-\frac{r_g}{r}\right)\frac{r^2}{r_g} \frac{d\chi_\mathcal{R}}{dr} \right]  
\end{equation}
and
\begin{equation}
\dot{p}_r^\mathcal{I} = \frac{\hbar^2}{2 c^2}\frac{\omega_\infty^2}{\left( 1-\frac{r_g}{r}\right)^2}\frac{r_g}{r^2}\left[\chi_\mathcal{I} - \left( 1-\frac{r_g}{r}\right)\frac{r^2}{r_g} \frac{d\chi_\mathcal{I}}{dr} \right] 
\end{equation}
By symmetry we fix the angular coordinate to the compact object equatorial plane,
$\theta=\pi/2$, which trivially sets the equatorial velocity $\dot{\theta}=0$ and equatorial momentum $p_\theta=0$. We also have vanishing rate of momentum change in the time and angular coordinates $\dot{p_t}=\dot{p}_\theta=\dot{p}_\phi=0$. The remaining non-vanishing equations of motion are:
\begin{equation}
p_t=-\frac{\hbar}{c}\omega_\infty
\end{equation}
and
\begin{equation}
p_r=\left[ \frac{\hbar^2}{c^2}\frac{\omega_\infty^2}{\left( 1-\frac{r_g}{r} \right)^2} \left( 1+\chi_\mathcal{R}+i \chi_\mathcal{I} \right) - \frac{p_\phi^2}{r^2}\frac{1}{\left(1-\frac{r_g}{r} \right)} \right]^\frac{1}{2},
\end{equation}
where the real and imaginary parts give $p_r^\mathcal{R}$ and $p_r^\mathcal{I}$, respectively. For the coordinate dynamics, we have
\begin{equation}
\dot{t}=Re\left\{\frac{\hbar}{c} \frac{\omega_\infty}{\left( 1-\frac{r_g}{r}\right)} \left( 1+\chi_\mathcal{R} + i \chi_\mathcal{I}\right) \right\},
\end{equation}
\begin{equation}
\dot{r}=Re\left\{ \left[ \frac{\hbar^2}{c^2} \omega_\infty^2 \left( 1+\chi_\mathcal{R} + i \chi_\mathcal{I} \right) - \frac{p_\phi^2}{r^2}\left(1-\frac{r_g}{r}\right)  \right]^\frac{1}{2} \right\}
\end{equation}
and
\begin{equation}
\dot{\phi}=\frac{p_\phi}{r^2}.
\label{eq:phiDot}
\end{equation}
Thus, we identify the constant with the angular momentum and set
\begin{equation}
    p_\phi = \text{constant} = L.
\end{equation}
We also identify the gravitational redshifted frequency
\begin{equation}
\omega(r)=\frac{\omega_\infty}{\sqrt{1-\frac{r_g}{r}}}.
\label{eq:SchwDopp}
\end{equation}
Input values are the starting coordinates $ct$, $r$, $\phi$, the asymptotic frequency $\omega_\infty$, the angular momentum $L$ as well as the mass $M$ and surface radius $R$ of the compact object that acts on the ray.

\section{A Dispersive Model for an Absorbing \\ Medium: Dusty Plasma}
\label{sec:dustyPlasma}

Dusty plasmas occur in many environments, in particular supernovae, neutron stars and in the accretion disks of black holes \cite{marklund}. Generally, cosmic dust has a complex index of refraction, explicitly showing the constant vacuum permittivity factor, 
\begin{equation}
 N=\sqrt{\mu \varepsilon_0(1+\chi_\mathcal{R} + i \chi_\mathcal{I})}   
\end{equation}
where we use the optical conductivity of dust $\sigma_d(\omega, x^a)$ to write the susceptibility as
\begin{equation}
\chi_\mathcal{I}(\omega, x^a)=\varepsilon_\mathcal{I}(\omega, x^a)=\frac{\sigma_d(\omega, x^a)}{ \varepsilon_0 \omega}.
\label{eq:opticalConductivity}
\end{equation}
Non-magnetic dust has $\mu=1$, and non-conducting dust has $\sigma_d=0$. Non-magnetic cold plasma with dielectric dust is described using the Drude model, 
\begin{equation}
    N^2 = 1 - \frac{\omega_p^2}{\omega^2 + i \frac{\omega}{\tau_d} }
    \label{eq:n2complex}
\end{equation}
with the plasma frequency 
\begin{equation}
\omega_p^2(r) = \frac{q^2 N_p(r) }{\varepsilon_0 m},
\label{eq:plasmaFreq}
\end{equation}
and $\tau_d$ is the average time between collisions of electrons with dust grains within the material, which can be given as a collision frequency $\omega_d=1/\tau_d$. The absorption process occurs due to the interactions of the colliding electrons within the dielectric. Electron collisions transfer the incident energy from an electromagnetic wave to the particles that the electrons interact with. Despite its origins in condensed matter physics, the Drude model has found applications in astrophysics for describing dust scattering throughout the electromagnetic spectrum (ie, from silicate and graphite grains in the $0.50$ - $2.00$ keV X-ray band \cite{drudeXRay}). At high energies we expect the refractive index to be nearly equivalent to the vacuum value $\chi_\mathcal{R} \approx 0$. However, for an X-ray photon a dust grain can act as both an absorber and scatterer of radiation. Thus, the imaginary part of the index of refraction is significant for energy transport \cite{draine2021}.

We seek the real and imaginary components of the susceptibility. Multiplying by the complex conjugate of the denominator gives 
\begin{equation}
    N^2(r) = 1 - \frac{\omega_p^2(r)}{\omega^2+\omega_d^2} + i \left( \frac{\omega_d}{\omega} \right)\frac{\omega_p^2(r)}{\omega^2+\omega_d^2}
    \label{eq:n2DrudeFrequency}
\end{equation}
We identify by inspection with eq. \ref{eq:complexIndRef},
\begin{equation}
    \chi_\mathcal{R} = - \frac{\omega_p^2}{\omega^2+\omega_d^2}
    \label{eq:chiRTau}
\end{equation}
\begin{equation}
    \chi_\mathcal{I} = \left( \frac{\omega_d}{\omega} \right)\frac{\omega_p^2}{\omega^2 + \omega_d^2}.
    \label{eq:chiID}
\end{equation}
For our calculations, we treat $\omega_d$ as a constant. 
Eq. \ref{eq:chiID} has the same form as eq. \ref{eq:opticalConductivity} which allows the optical conductivity to be easily identified. When the collision frequency is small $\omega_d \rightarrow 0$, the cold plasma dispersion relation is recovered: 
\begin{equation}
    \lim_{\tau_d \rightarrow \infty} \chi_\mathcal{R}(\omega, \tau_d) = - \frac{\omega_p^2}{\omega^2}
    \label{eq:lim1}
\end{equation}
\begin{equation}
    \lim_{\tau_d \rightarrow \infty} \chi_\mathcal{I}(\omega, \tau_d) = 0
    \label{eq:lim2}
\end{equation}
In the small $\omega_d$ limit, the average time $\tau_d$ between electron collisions and dust grains is large, so the dust is diffuse and the dispersion relation reduces to the cold-plasma case. As we decrease the collision frequency $\omega_d$, we increase the refractive effect of the dielectric medium. In the opposite limit as the collision timescale becomes small $\tau_d \rightarrow 0$ such that $\omega_d \rightarrow \infty$, the vacuum behaviour is recovered. This limiting behaviour shows that $\chi_\mathcal{I}$ has a maximum at some intermediate value of $\omega_d$. Both of these limits have been well-studied in the gravitational lensing literature \cite{olegReview}. For a constant $\omega_d$, we are describing dielectric dust and cold plasma that homogeneously fills the spacetime. The density of the medium decreases with radius following the plasma frequency. The constant $\omega_d$ is essentially an interpolation parameter that smoothly interpolates between the vacuum solution for large $\omega_d$, and the cold plasma solution for small $\omega_d$.

Evaluating the index of refraction $N^2(r)$ in eq. \ref{eq:n2DrudeFrequency} requires a specific form for the plasma frequency. We adopt a power-law form for the plasma electron density
\begin{equation}
    N_p(r)=N_{p0}\left( \frac{R}{r}\right)^h
\end{equation}
in terms of the power-index $h>0$. As defined in eq. \ref{eq:plasmaFreq}, we have  
\begin{equation}
    \omega_p^2(r) = K_p\left(\frac{R}{r} \right)^h
    \label{eq:plasmaPowerLaw}
\end{equation}
with the constant 
\begin{equation}
    K_p = \frac{q^2 N_{p0}}{\varepsilon_0 m}.
\end{equation} 
In this expression we use $q$ as the fundamental electron charge, $m$ the electron mass and $R$ the stellar radius and $N_{p0}$ is the maximum plasma density.

Throughout the remainder of this work we will treat the Drude collision frequency as a constant for simplicity. This leads to the plasma frequency controlling the spatial behaviour of both $\chi_\mathcal{R}$ and $\chi_\mathcal{I}$. However, we do not expect this simplifying assumption to hold physically as the dust density is not expected to remain constant throughout spacetime. Generally the collision timescale and dust density are inversely proportional to one another. In Schwarzschild spacetime near the central mass (small $r$), we expect the dust to be in a high density state and therefore have a high collision frequency $\omega_d$. Conversely, at large $r$ we expect low density, and therefore a low Drude collision frequency $\omega_d$. The limits considered in eq \ref{eq:lim1} and \ref{eq:lim2} demonstrate this inverse behaviour at large $r$. In fact, the timescale itself is built up of the sum of several separate effects. The electrons interact with electromagnetic waves and dust particles on their own unique timescales. Thus, modulo a factor of $2\pi$ which is absorbed into $\tau$, we write
\begin{equation}
    \frac{1}{\tau_d} = \omega_d = \omega_{EM} + \omega_{dust} 
\end{equation}
where $\omega_{EM}$ and $\omega_{dust}$ are the interaction frequencies of electrons in the plasma with electromagnetic waves and dust particles, respectively. The timescale is the sum of the interaction frequencies of the electrons with all other particles in the medium. Due to the small interaction frequency with protons in the plasma, we are safe to neglect this extra contribution.

\section{Ray-Tracing in a Strongly Absorbing \\ Medium}
\label{sec:rayTracingStrong}

\begin{figure*}
\centering
\includegraphics[viewport=25 20 390 200, clip=true, scale=1.05]{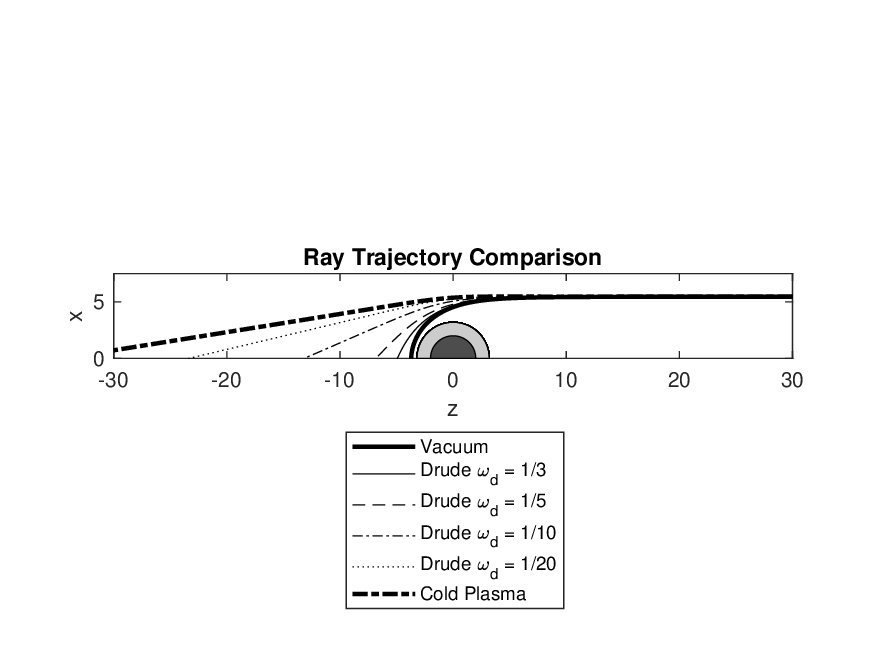}
\caption{Schwarzschild compact object surrounded by a Drude medium. In the limit of low collision frequency $\omega_d$ the cold plasma behaviour is recovered and for high collision frequency the Drude medium reproduces the Schwarzschild vacuum behaviour. We used a constant relaxation time and a power-law plasma density with $k_p=1$ and $h=3$. The compact object is given by $M=1$, $R=3.2$ in dimensionless code units.}
\label{fig:tauCompare}
\end{figure*}

As a demonstration that the Drude medium interpolates between cold plasma and vacuum cases, consider the scenario depicted in fig. \ref{fig:tauCompare}. In this figure we show the result of a ray tracing calculation with various realizations of a constant Drude collision frequency $\omega_d$. For the plasma frequency, we use a power-law with $K_p=1$ and power-index $h=3$. This figure shows that the Drude model reproduces the black hole surrounded by cold plasma and Schwarzschild vacuum lensing behaviour for the appropriate limit of the Drude collision frequency $\omega_d$. To generate this figure we used a compact object with mass $M=1$, and radius $R=3.2$ in scaled units. The physical surface of the object is colored light gray and the event horizon is colored dark gray to demonstrate the physical scale of object we are considering. 

\begin{figure*}
\centering
\includegraphics[viewport=20 20 400 300, clip=true, scale=1.05]{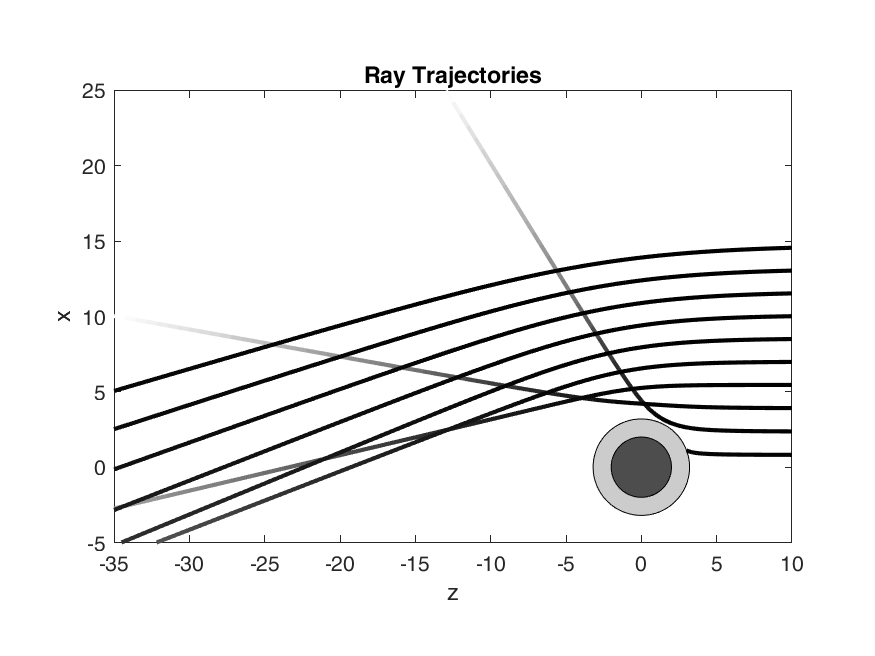}
\caption{An example of refractive effects around a Schwarzschild compact object surrounded by Drude medium. Shading of the rays indicates absorption along a trajectory. System parameters are $M=1$, $R=3.2$, $k_p=1$, $h=3$, $\omega_d=1/15$. The ray frequency was chosen to be $1 \%$ greater than the plasma frequency on the stellar surface to maximize the medium effects. The rays show a diversity of behaviours including both converging and diverging lens effects.}
\label{fig:wdReal}
\end{figure*}

The absorption that a ray experiences travelling through a Drude medium is shown in fig. \ref{fig:wdReal} which demonstrates ray tracing around a compact object surrounded by a spherically symmetric Drude medium. We assume a power-law in plasma density with $K_p=1$ and $h=3$ and Drude collision frequency $\omega_d=1/15$. Shading along the rays represents the absorption, with the most dramatically affected rays sampling the high plasma density region near the origin. Shading was calculated by numerically integrating the imaginary part of the eikonal (eq. \ref{eq:Seik}) along each ray path. We determine the ray paths using the real valued Hamiltonian procedure detailed in Section \ref{sec:opacity}.

\section{Weak Field Limit with Weak Absorption: Microlensing Behaviour}
\label{sec:ML}

We will use the weak field limit of the metric to find the deflection angle for a point mass lens, surrounded by an absorbing, dispersive dielectric medium. In this case, the lensing behaviour occurs when spacetime is only gently curved. The weak-field limit of the metric is 
\begin{equation}
g^{ik}=\eta^{ik}-h^{ik}    
\end{equation}
where $\eta^{ik}$ is the Minkowski metric. In addition to the gravitational field, we also expect the effect of absorption to be similarly weak due to the scale of the deflecting system. Following \cite{chenKantowski1}, we assume a weakly absorbing medium, which is written
\begin{equation}
N(r) = n(r) + i\kappa(r) \approx n(r)(1+\lambdabar \bar{\kappa})
\end{equation}
with $\kappa / n = \lambdabar\bar{\kappa} + \mathcal{O}(\lambdabar^2)$. This has consequences on the equations of motion. We return to eq. \ref{eq:complexMom1} and assume the imaginary part is weak, 
\begin{equation}
    p_r = p_r^\mathcal{R} + i \lambdabar p_r^\mathcal{I}
\end{equation}
which we write in terms of a small coefficient, which we associate with the geometric optics expansion parameter. Then, since the square of the radial momentum enters the Hamiltonian, the correction to the equations of motion from the medium reaction is to second order in the expansion parameter. In the weak absorption limit, it is safe to use the deflection purely due to the real part of the index of refraction $n(r)$. Thus, the optical metric is also real to first order, leading to a first-order real eikonal \cite{chenKantowski1}. In this case the absorption is weak since it occurs over many wavelengths, which transfers the absorption term from the eikonal to the wave amplitude. In the weak-field limit the absorption occurs over many wavelengths. As shown in \cite{chenThesis}, with a real eikonal the amplitude transport equation (eq. \ref{eq:propODE1}) acquires an extra extinction term which is given by
\begin{equation}
    \tau = 2\int_{\ell_0}^\ell \bar{\kappa} n^2 \frac{(u^ak_a)^2}{c^2} d \ell. 
    \label{eq:opticalDepthML2}
\end{equation} 
Simplifying using eqs. \ref{eq:opticsImag} and \ref{eq:reducedWavelength} gives the optical depth in the weak-deflection limit,
\begin{equation}
    \tau = \int_{\ell_0}^\ell \chi_\mathcal{I} \frac{\omega}{c} d \ell. 
    \label{eq:opticalDepthML}
\end{equation}

The microlensing calculation proceeds generally following the approach outlined in the work of Tsupko \& Bisnovatyi-Kogan \cite{olegReview, olegShadow, olegPlasmaTimeDelay, olegRadioLens, olegGeneral, oleg_volker, olegRadioSpectrometer}. We will review the general features of the calculation here. Since the gravitational field is weak, we may consider the null approximation for the light ray in which the deflected path is broken into two straight-line segments (see fig. 1 in \cite{oleg2} or figs. 3 and 4 in \cite{narayan} for example). The gravitating mass is physically small with respect to the scale of the overall light path, and the region of interaction between the light ray and gravitating body is correspondingly small. The thin lens approximation allows the effect of the lens to be treated as if the entire deflection of the light ray occurs as it passes through the plane containing the compact object $M$ perpendicular to the ray path. Orienting the undeflected ray along the $z$-axis, we can describe the deflection angle in terms of the two components of the lens plane perpendicular to the ray $(x,y)$. Specifying positions on the lens plane is simplified further by assuming spherical symmetry and giving the deflection angle in terms of the impact parameter.

Let our coordinate system be fixed such that the light ray is launched along the $z$-axis. We write the covariant momentum $4$-vector as
\begin{equation}
    p_a=\left( -\frac{\hbar \omega_\infty}{c}, 0, 0, \frac{n_\infty \hbar \omega_\infty}{c}\right).
    \label{eq:momNullCov}
\end{equation}
Since the photon is travelling in the $z$-direction, the unit vector along the undeflected ray path is particularly simple, $\hat{e}^\alpha = \hat{e}_\alpha = (0,0,1)$, and the spatial components of the momentum $3$-vector are in terms of the unit vector along the unperturbed trajectory, 
\begin{equation}
    p_\alpha = p\hat{e}_\alpha = |p_z|\hat{e}_\alpha = \frac{n_\infty \hbar \omega_\infty}{c}\hat{e}_\alpha.
    \label{eq:pAlphaVecDefl}
\end{equation}
A spherically symmetric lens allows us to change from a two-dimensional description of the lens plane in $(x,y)$ coordinates to a single dimension using the impact parameter $b$. We can now write $r=\sqrt{b^2 + z^2}$ by assuming the ray pierces the lens plane a radial distance $b$ from the lens center. Now the planar components $\alpha=1,2$ are associated with the radial direction. The redshift factors are safe to be neglected since we are assuming the weak-field metric, so the frequency of the ray remains approximately constant on its journey through the lens $\omega(x^a) \approx \omega_\infty$.

From here, we follow the steps outlined by Bisnovatyi-Kogan and Tsupko \cite{olegRadioSpectrometer}, with a general index of refraction. The deflection angle is expressed as an integral over a sum of terms. Using frequency as constant, we have the general expression 
\begin{equation}
    \hat{\alpha}_b = \frac{1}{2}\int_{-\infty}^{+\infty}\frac{b}{r} \left[ \frac{d h_{zz}}{d r} + \frac{1}{n_\infty^2}\frac{d h_{tt}}{d r} + \frac{1}{n_\infty^2  }\frac{d \chi_\mathcal{R}}{d r}  \right]dz.
    \label{eq:MLIntegral}
\end{equation}
Progress requires a particularly useful integral identity \cite{olegRadioSpectrometer, integrals},
\begin{equation}
    \int_0^{+\infty} \frac{dz}{(z^2 + b^2)^{\frac{h}{2}+1}} = \frac{1}{hb^{h+1}}\frac{\sqrt{\pi}\Gamma\left( \frac{h}{2} + \frac{1}{2} \right)}{\Gamma\left(\frac{h}{2} \right)}.
    \label{eq:integralIdentity}
\end{equation} 

We use the weak field metric components $h_{tt} = r_g/r$, $h_{zz} = r_g/r \cos^2\theta$ with $\cos \theta=z/(b^2+z^2)^\frac{1}{2}$. The simplest case is when the real susceptibility is constant $\chi_\mathcal{R}=\chi_{\mathcal{R}0}$ throughout space. In that case, the term containing the derivative vanishes, and we are left with 
\begin{equation}
    \hat{\alpha}_b = - \frac{r_g}{b} \left( 1+\frac{1}{1+\chi_{\mathcal{R}0}} \right)
\label{eq:constML}
\end{equation} 
as found in \cite{olegRadioLens}. Assuming the plasma density is constant, we evaluate the optical depth of the medium using eq. \ref{eq:opticalDepthML}. Evaluating the integral over all space leads to divergence, which represents the photon being completely absorbed after travelling an infinite distance through a medium of finite absorption. Therefore, we write the optical depth using \ref{eq:opticalDepthML} as
\begin{equation}
    \tau = \chi_{\mathcal{I}0} \frac{\omega_\infty}{c} z
    \label{eq:tauConstant}
\end{equation}
with $\chi_{\mathcal{I}0}$ constant throughout all space and with $z$ the distance from source to lens. Formally, the evaluation of the integral in eq. \ref{eq:opticalDepthML} diverges for infinite limits with a constant $\chi_{\mathcal{I}0}$, and thus we write the optical depth as a function of distance along the undeflected path from source to observer $z$. This ray suffers exponential loss along its trajectory. Such a lensing event would be unobservable at sufficiently great distances due to the absorption of the intervening medium between source and observer.   

In fact, the previous case is not realistic due to the constant index of refraction which we generally expect to vary through space. We set $n_\infty=1$ as the boundary condition at infinity. This allows us to rewrite eq. \ref{eq:MLIntegral}, the deflection angle integral, 
\begin{equation}
    \hat{\alpha}_b = -\frac{r_g}{b} - \frac{r_g b}{2}\int_{-\infty}^{+\infty} \frac{1}{r^3}dz 
+ \frac{b}{2}\int_{-\infty}^{+\infty} \frac{1}{r}\frac{d\chi_\mathcal{R}}{dr}dz.
\label{eq:stepMicroLens}
\end{equation}
Consider the first integral term, which is evaluated using the identity from eq. \ref{eq:integralIdentity}, 
\begin{equation}
    \int_{-\infty}^{+\infty} \frac{1}{(b^2+z^2)^\frac{3}{2}}dz = \frac{2}{b^2}, 
\end{equation}
further simplifying the deflection angle formula 
\begin{equation}
    \hat{\alpha}_b = -\frac{2 r_g}{b} + \frac{b}{2}\int_{-\infty}^{+\infty} \frac{1}{r}\frac{d\chi_\mathcal{R}}{dr}dz.
\label{eq:stepMicroLens2}
\end{equation}
In the absence of any medium, we recover the vacuum Schwarzschild lensing behaviour. Using the power-law form for the plasma frequency (eq. \ref{eq:plasmaPowerLaw}), we find the deflection angle for the Drude medium
\begin{equation}
    \hat{\alpha}_b = -\frac{2 r_g}{b} + \frac{1}{b^h}\frac{K_p R^h}{(\omega^2 + \omega_d^2)}\frac{\sqrt{\pi}\Gamma\left( \frac{h}{2}+ \frac{1}{2}\right)}{\Gamma\left( \frac{h}{2}\right)}
    \label{eq:deflAngleA}
\end{equation}
which reproduces the cold plasma deflection angle in the limit $\omega_d \rightarrow 0$ \cite{olegRadioSpectrometer}. In deriving this formula we have used the properties of the gamma function, $\Gamma(1)=1$ and $\Gamma(1/2)=\sqrt{\pi}$. 

The optical depth of the Drude medium is evaluated using eq. \ref{eq:opticalDepthML}. The integral along the line of sight requires using the integral identity eq. \ref{eq:integralIdentity} with $H=h+2$, which gives the new relationship, 
\begin{equation}
    \int_0^\infty \frac{dz}{(b^2 + z^2)^\frac{H}{2} } = \frac{1}{(H-2)}\frac{1}{b^{(H-1)}} \frac{\sqrt{\pi} \Gamma\left( \frac{H}{2}-\frac{1}{2} \right)}{\Gamma\left( \frac{H}{2}-1 \right)}.
    \label{eq:integralIdentity2}
\end{equation}
Using the imaginary part of the susceptibility $\chi_\mathcal{I}$ (eq. \ref{eq:chiID}), we find
\begin{equation}
    \tau = \frac{ \omega_d}{c}\frac{K_p R^h}{(\omega^2 + \omega_d^2)} \int_{-\infty}^{+\infty} \frac{1}{(b^2 + z^2)^\frac{h}{2}} dz
\end{equation}
Using the second form of the integral identity (eq. \ref{eq:integralIdentity2} with the label $h$ in place of $H$), we evaluate this expression and find
\begin{equation}
    \tau = \frac{2}{(h-1)}\frac{ K_p R^h }{b^{h-1}} \frac{\omega_d }{c}\frac{1}{(\omega^2 + \omega_d^2)}\frac{ \sqrt{\pi}\Gamma\left( \frac{h}{2}+\frac{1}{2}\right)}{\Gamma\left( \frac{h}{2}\right)} 
    \label{eq:DrudeOpticalDepth1}
\end{equation}
which is valid for $h > 1$. When we consider the $h=1$ case, we find the integral diverges as
\begin{equation}
    \tau = 2 \frac{K_p R}{c} \frac{ \omega_d }{(\omega^2 + \omega_d^2)}  \ln \left| \sec \Theta + \tan \Theta \right| 
    \label{eq:DrudeOpticalDepth2}
\end{equation}
evaluated in the limit as $\Theta \rightarrow \pi/2$. Similar to the case of absorption for a constant complex susceptibility (eq. \ref{eq:tauConstant}), the divergence of the integral represents the total extinction of the photon due to the distance travelled through an absorbing medium, which is formally infinite. 

For $h>1$ the power-law density vanishes at an infinite distance from the origin. The density of the medium varies substantially enough from source to observer that the integral does not diverge even for infinite separation. With this expression, we can evaluate the optical depth over the plane of the sky, similar to the deflection angle. The deflection angle (eq. \ref{eq:deflAngleA}) and the expressions for the optical depth (eqs. \ref{eq:DrudeOpticalDepth1} \& \ref{eq:DrudeOpticalDepth2}) fully characterize the weak-field microlensing behaviour of the Drude medium with a power-law density in the Schwarzschild spacetime.

The thin lens equation that describes the coordinate transformation from source plane to lens plane is \cite{schneider, narayan}
\begin{equation}
    \beta = \theta - \alpha(\theta)
\end{equation}
with the deflection angle 
\begin{equation}
    \alpha\left( \theta \right) = -\frac{D_{ds}}{D_s} \hat{\alpha}\left(b\right)
    \label{eq:scaledDeflAngle}
\end{equation}
using the relationship $b=D_d \theta$. In the usual thin-lens formalism, the sign is absorbed into the thin lens equation, and the vacuum deflection of a point source (the first term in eq \ref{eq:deflAngleA}) is taken as positive. This is reflected in the sign used in eq \ref{eq:scaledDeflAngle}. Let us also define the Einstein radius 
\begin{equation}
     \theta_E^2 = \frac{4 G M}{c^2  }\frac{D_{ds}}{D_d D_s}
\end{equation}
and the Drude scale
\begin{equation}
   \theta_D^{h+1}=  K_p \frac{D_{ds}}{D_s} \frac{R^h}{D_d^h} \frac{\sqrt{\pi} \Gamma\left(\frac{h}{2}+\frac{1}{2} \right)}{\Gamma\left( \frac{h}{2} \right)}.
\end{equation}
With the weak-field lensing sign convention the deflection angle becomes
\begin{equation}
    \alpha(\theta) = \frac{\theta_E^2}{\theta} - \frac{\theta_D^{h+1}}{\theta^{h}}\frac{1}{(\omega^2 +\omega_d^2)}
\end{equation}
and the optical depth ($h>1$) over the lens plane is 
\begin{equation}
    \tau(\theta) = \frac{2}{(h-1)}\frac{D_d D_s}{D_{ds}} \frac{\omega_d}{c}\frac{1}{(\omega^2 + \omega_d^2)} \frac{\theta_D^{h+1} }{\theta^{h-1}}.
    \label{eq:optDepFin}
\end{equation}

\begin{figure*}
\centering
\includegraphics[viewport=20 50 425 250, clip=true, scale=0.95]{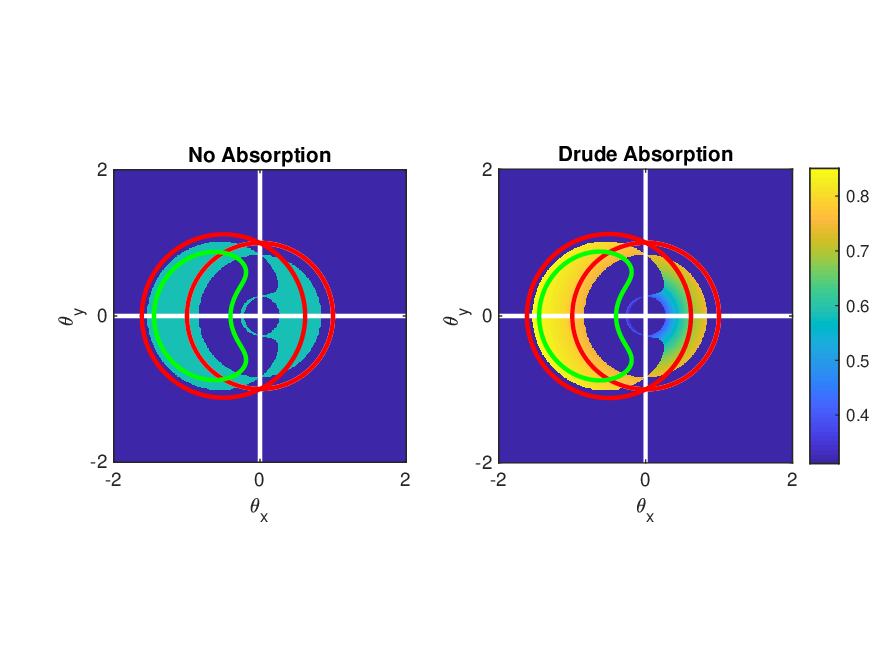}
\caption{Lensed view of extended background source, a disk of uniform brightness $I_0=1$. Left panel, no absorption. Right panel includes Drude absorption. The central ring-like image is eliminated due to the Drude absorption. White crosshair lines mark the location of the lens center. Red contours show the Schwarzschild vacuum image, green contours outline the Schwarzschild and cold plasma image. The lens has $\theta_E=1$ and the Drude parameters are $\theta_D=0.50$, power-law constants $k_p=1$ $h=2$, collision frequency $\omega_d=0.50$ and asymptotic frequency $\omega_\infty=0.50$. Calculations were performed in normalized units.}
\label{fig:ML1}
\end{figure*}

\begin{figure*}
\centering
\includegraphics[viewport=0 0 375 350, clip=true, scale=0.80]{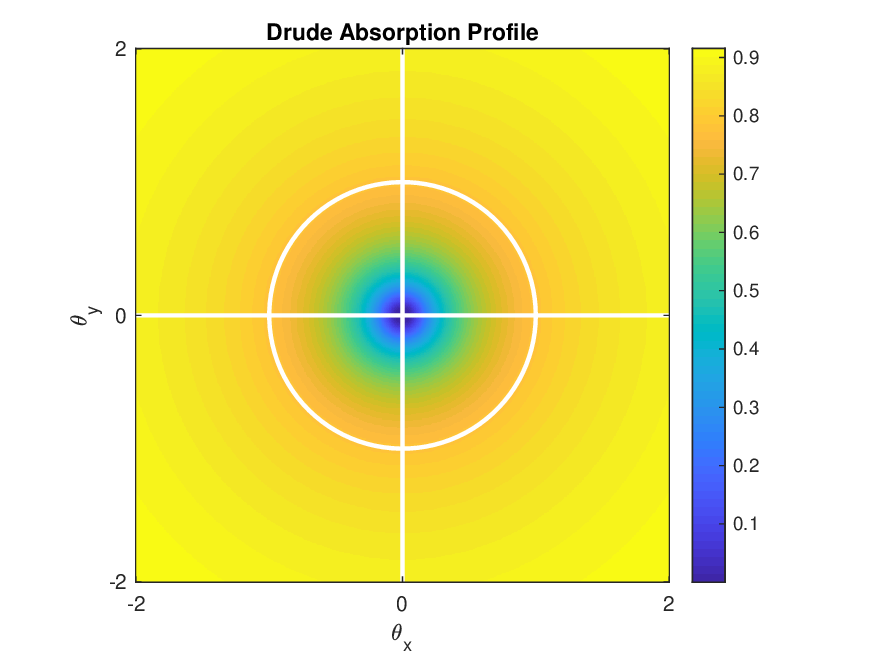}
\caption{Drude absorption factor $\exp[-\tau(\theta)]$ over the lens plane. The crosshair and Einstein ring are marked as thin white lines. The parameters correspond to the lens configuration in fig \ref{fig:ML1}.}
\label{fig:ML2}
\end{figure*}

 We demonstrate the effect of the Drude absorption factor in figs \ref{fig:ML1} and \ref{fig:ML2}. In fig \ref{fig:ML1} we show an example lensed image of an extended source. We use a disk with radius $0.5$ as the source. We assume even illumination $I_0=1$ across the face of the disk. The lens is a point mass with $\theta_E=1.00$, surrounded by an absorbing Drude medium described by the constant $\theta_D=0.50$. We consider observations at asymptotic frequency $\omega_\infty=0.50$ and collision frequency $\omega_d=0.50$ in normalized units. We set the plasma constants $K_p=1$ and $h=2$. We arbitrarily set the distance factor in the optical depth to unity (ie, taking the lens halfway between source and observer such that $D_d=D_{ds}=D_s/2$). The red contours show the boundary of the Schwarzschild vacuum image, and the green contour shows the Schwarzschild plus cold plasma image. The white crosshairs indicate the location of the lens center. The vacuum Einstein ring is obscured by the image contours. The left panel shows the Schwarzschild plus Drude image without the absorption applied, and the right panel shows the same image configuration with the absorption effect included. In this example, the absorption reduces the intensity of the interior image regions by a substantial amount (to $\sim 30\%$ of the vacuum case). The absorption at the lens center effectively removes the center ring-image. The absorption profile is shown in fig \ref{fig:ML2}, and demonstrates that absorption can have a substantial impact on the appearance of extended sources observed through gravitational lenses surrounded by absorbing material.

Next, let us investigate the effect of absorption on the light curve of a lensed point-source. We will make use of the same lens parameters as discussed above for the extended disk source. We choose an impact parameter $b$, and use a constant source brightness $I_0=1$. The location of the lensed images on the lens plane is used to evaluate the optical depth at that location on the observers sky and determine the amount of absorption each individual image experiences. This is calculated using eq \ref{eq:optDepFin} for each solution of the thin lens equation $\tau_a=\tau_D(\theta_a)$, along with the lensed magnification $\mu_a=\mu(\theta_a)$. The total magnification for a point source at a given impact parameter is the sum of the product of these two factors, 
\begin{equation}
    \mu_T(b) = \sum_{a=1}^N \exp(-\tau_a) |\mu_a|.
    \label{eq:mut}
\end{equation}
Since the Schwarzschild-Drude medium lens is spherically symmetric, the magnification is straightforward to calculate,
\begin{equation}
    \mu(\theta) = \frac{\theta}{\beta} \frac{d\theta}{d\beta} 
\end{equation}
with 
\begin{equation}
    \frac{\beta}{\theta} = 1-\frac{\theta_E^2}{\theta^2} + \frac{\theta_D^{h+1}}{\theta^{h+1}}\frac{1}{(\omega^2 + \omega_d^2)}
    \label{eq:mag1}
\end{equation}
and
\begin{equation}
    \frac{d \beta}{d \theta} = 1 + \frac{\theta_E^2}{\theta^2} -h \frac{\theta_D^{h+1}}{\theta^{h+1}}\frac{1}{(\omega^2 + \omega_d^2)}.
\label{eq:magDeriv}
\end{equation}
Additionally, the lens critical curves can also be found from where these expressions vanish. The first of these conditions (eq \ref{eq:mag1}) is written in the form of a polynomial of order that depends on $h$,
\begin{equation}
    \theta^{h+1} - \theta_E^2 \theta^{h-1} + \frac{\theta_D^{h+1}}{(\omega^2 + \omega_d^2)}=0.
\end{equation}
The solution of this polynomial yields the tangential critical curves. The derivative expression eq. \ref{eq:magDeriv} yields the radial critical curves \cite{rogersEr2018},
\begin{equation}
\theta^{h+1} + \theta_E^2 \theta^{h-1} - h\frac{\theta_D^{h+1}}{(\omega^2 + \omega_d^2)} =0.
\end{equation}

\begin{figure*}
\centering
\includegraphics[viewport=50 0 375 325, clip=true, scale=1.0]{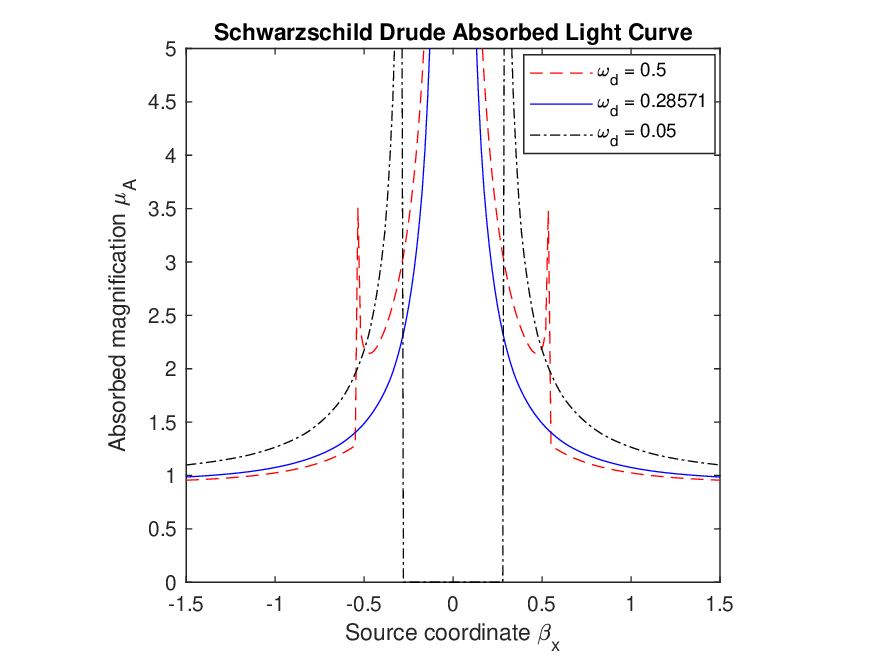}
\caption{Light curve for a point source observed through the Schwarzschild-Drude lens. The $y$-component of the impact parameter was set to the constant value $b_y=0.01$. To generate the light curves we used the following parameters: $\theta_E=1$, $\theta_D=0.50$, $\omega_\infty=0.50$, $h=2$ for a variety of collision frequency values $\omega_d$, selected to provide three distinct morphologies of light curve.}
\label{fig:ML_lightcurve1}
\end{figure*}

We show the point source light curve in fig \ref{fig:ML_lightcurve1}. We use the impact parameter $\beta_y=0.01$ and allow the $x$-component of the trajectory to vary $-1.5 \leq \beta_x \leq 1.5$. We used Einstein ring radius $\theta_E=1$, Drude scale $\theta_D=0.50$, asymptotic frequency $\omega_\infty=0.50$ and plasma power-index $h=2$. We use a variety of collision frequency values scaled by the asymptotic frequency $\omega_d=\omega_\infty/f$, selected arbitrarily to display light curves with three distinct morphologies: $f = 1$ (red dashed curve), $f = 1.75$ (blue solid curve), $f = 10$ (black dash-dotted curve).

The light curves do not show a qualitative difference compared with the non-absorbed case in terms of the morphology of the curves. However, the absorption in this example reduces the overall magnification near the central part of the lens by $\approx 15 \%$. The absorption near the lens center has a sizable effect on the observed magnification by reducing the intensity. The $f=10$ case (black curve) also shows a large exclusion region where no images are formed, analogous to the phenomenon that occurs with the cold plasma lens \cite{rogersEr2018}.

\section{The Linear Kramers-Kronig Relationship}
\label{sec:KK}

The real and imaginary parts of the complex index of refraction are not independent or arbitrary. In fact, they are intimately linked through the enforcement of causality by the Kramers-Kronig relationship, which dictates the behaviour of realistic dielectric media. 

The Drude model in eq. \ref{eq:n2complex} is a single pole model, using only one relaxation constant $\tau_d$. The Linear Kramers-Kronig relation (Lin-KK; \cite{linKK1, linKK2, linKK3}) uses the Drude model as a physically realistic basis function for a collection of independent relaxation process. For a given real index of refraction, this multi-pole approach fits a linear model over a given frequency range and produces the corresponding physically-linked component.

Consider a series of terms $1, .., M_p$ that describe $M_p$ individual relaxation constants $\tau_{d1}$, $\tau_{d2}$, .., $\tau_{dM_p}$. Let the relaxation constants be logarithmically distributed between the minimum $\tau_{min} = \tau_{d1} = \omega_{max}^{-1}$ and maximum $\tau_{max}= \tau_{dM_p}=\omega_{min}^{-1}$. Then we will have a sequence of these relaxation constants such that  
\begin{equation}
    \tau_b = 10^{\left[ \log \left(\tau_{min}\right) + \frac{b-1}{M_p-1} \log \left( \frac{\tau_{max}}{\tau_{min}} \right) \right]}
\end{equation}
with $b=2, .., M_p$. Let us define a set of real $M_p$ parameters for the index of refraction $P_1$, $P_2$, $P_3$, .., $P_{M_P}$. We write the square of the refractive index
\begin{equation}
    N^2(\omega) = P_1 - \sum_{a=2}^{M_p} \frac{P_a}{\omega^2 + i \frac{\omega}{\tau_a}}.
    \label{eq:n2KK1}
\end{equation}

Let us also expand this function to explicitly show the real and imaginary parts. Using the complex conjugate of the denominator, we find
\begin{equation}
    N^2(\omega)= P_1 - \sum_{a=2}^{M_p}P_a\left[ \frac{1}{(\omega^2 +\omega_d^2)} -i\left( \frac{\omega_d}{\omega} \right) \frac{1}{(\omega^2 + \omega_d^2)} \right].
    \label{eq:drudeKK}
\end{equation}
This expansion uses the Drude model as a basis function to fit an index of refraction and generate the missing component. Either real or imaginary components can be the fit function. Due to the ubiquity of complex spectra, this basic method has been applied in many areas, including the analysis of impedance data \cite{linKK1, linKK_esteban, linKK_agarwal1, linKKBackup3} and in measurements of electrochemical immittance \cite{linKK2}. With one term we recover the Drude model in eq. \ref{eq:drudeKK} if we take the parameter values as $P_1=1$ and $P_2 = \omega_p^2$. With this model and a proposed dispersion-dependent real function (representing $\chi_\mathcal{R}$ or $\chi_\mathcal{I}$), we can find a corresponding self-consistent, physically plausible partner to complete the complex extension of the full index of refraction. 

Note that the index of refraction defined by the linear Kramers-Kronig expansion does not include any spatial dependence. When the index of refraction is more complicated, such as when the plasma frequency is not a constant with respect to the coordinates, the parameters must inherit the spatial dependence by assuming a functional form for the variation of the fit parameters (a power-law in our examples). Throughout our work we have used $P_1=1$, but we leave eq \ref{eq:drudeKK} totally general and include it as an adjustable parameter, giving a total of $M_p$ fit parameters.

\section{Discussion}
\label{sec:disc}

Throughout this work, we have not had physical motivation to choose anything other than real-valued constants of motion. We have simply set the imaginary parts of the integration constants to vanish. This produces simple constraints for the imaginary part of the Hamiltonian: namely, a loss of radial momentum along a trajectory. The radial imaginary components of the eikonal are a consequence of the spherically symmetric complex index of refraction. In general we could include a tensor-valued index of refraction, which could induce momentum loss in both the radial and equatorial directions. If angular momentum were included as a complex variable, then a corresponding angular term would appear in the imaginary part of the eikonal, and rays would show a larger variety of more complicated dynamics. This possibility is outside the scope of this work, but would make an interesting follow-up study. 

We assumed a constant collision frequency throughout all spacetime. However, a more realistic model would be to assume a functional form for spatial dependence of the collision frequency $\omega_d=\omega_d(x^a)$. For example, the Drude model could be adapted to feature independently radially varying plasma and dust density distributions. A spherically-symmetric plasma density distribution has $\omega_p(r)$ (eq. \ref{eq:plasmaPowerLaw}), and we could imagine a spherically-symmetric dust collision frequency $\omega_d(r)$, also given by a power-law form for the dust profile,
\begin{equation}
    \omega_d(r)^2 = K_{d}\left( \frac{R}{r} \right)^{h_{d}}.
\end{equation}
We emphasize that $K_d$ and $h_d$ need not match the parameters of the plasma density $K_p$ and $h$ and could be selected independently. We speculate both plasma and dust distributions plausibly mirror one another in the Schwarzschild spacetime, but this assumption fails in the case of strong magnetic fields, in which we expect highly anisotropic distributions for both plasma and dust. This choice of parameterization, which includes spatial dependence, is more realistic. In general, power-laws as functional forms for density remain an arbitrary and physically-unmotivated but convenient assumption. Due to this arbitrariness our model serves as a useful toy example.

The model presented here, paired with a description of magnetic fields \cite{heyl1, magBH1, magBH2}, opens new opportunities for physical modeling due to dust anisotropy. In a previous paper \cite{rogers15} we described the visibility of hot spots \cite{NSPulseOrig, NSPulseCharged, NSPulseTurolla} on the surface of a neutron star when surrounded by a distribution of plasma. This refractive environment modified the pulse profiles sufficiently from their vacuum properties that the effect could possibly be observed in the neighborhood of $100$ MHz, with the strength of the effects growing at lower frequencies that are obscured by Earth-based observations. The pulse-profile calculation has been expanded on in a particularly illuminating and thorough analytical dissection by \cite{galloNSA}. It is well-known that dust is relevant to the appearance of X-ray pulse profiles \cite{Xray1, Xray2}. A detailed study would be useful to find the corresponding extinction affecting a hot spot on the surface of a dust-obscured compact object.

% Isolated diverging lens objects comprised of some combination of dielectric dust and plasma could also be a population of interesting targets for producing extreme scattering events given the ubiquity of plasma lensing models \cite{ESE1, ESE2, ErESE}. Additionally, diverging lens models are often discussed in relation to fast radio bursts \cite{erFRB, erFRB2}. Isolated structures of Drude-type media both refracting and absorbing photons may also produce complicated lensing phenomena that modify the brightness of background objects beyond conventional lensing magnification calculations.

Finally, details of the black hole shadow \cite{olegShadow, perlick24} are of high interest now that the Event Horizon Telescope has imaged the photon sphere of black holes directly \cite{EHT}. A further worthwhile extension of this work could include a study of a black hole shadow including intensity modification effects due to dust. In addition, recent work has focused on more realistic accretion flows in both Schwarzschild and Kerr metrics \cite{oleg24}. The addition of more realistic fluid flows would also be a valuable expansion of this work. 

\section{Conclusions}
\label{sec:conclusions}

We studied Synge's medium equation with a complex index of refraction, leading to the interpretation of the complex Hamiltonian. The complex Hamiltonian describes the dynamics of both the phase and amplitude of an electromagnetic wave travelling along a geometric ray. We have provided a physical interpretation of the real and imaginary parts of the eikonal, illustrated the dynamics using ray-tracing with a complex index of refraction in the presence of strong absorption, and generalized the model to the weak-field deriving the optical depth for a Drude medium in the Schwarzschild background. We are able to describe both the refractive and absorptive properties of a material medium on curved space time and the effect on electromagnetic radiation, extending the practical range of Synge's medium equation and extended the results of Chen \& Kantowski. We find that the absorption due to dust distributions that drop off more slowly than a power-law with index $h=1$ experience exponential absorption along their paths. For $h>1$ we obtain analytical results for the absorption due to a projected distribution of Drude material.

In terms of observable results, we have shown the effect of dust scattering in the Drude model acts as a parameter to interpolate between vaccuum and cold plasma cases. The cold plasma effects begin to become observable around the $1$ GHz range, but depending on the plasma density we expect the most prominent display at low frequencies on the order of $100$ MHz and below. In contrast, the dust absorption of the Drude model has been applied in high frequency regimes such as the X-ray band. 

Our examples demonstrate that dielectric absorption can have a substantial impact on the appearance of sources observed through gravitational lensing. This study leads to a variety of problems that we have suggested for follow up work. In Section \ref{sec:disc} we describe open questions and potential future work to further extend this topic in interesting and novel directions. 

\section*{Acknowledgements}
This paper is dedicated to the memory of Ian D. Cameron, Director of Lockhart Planetarium, astronomer, instructor, colleague, co-author, collaborator, coffee buddy, founding member of the Manitoba Magic and Latin Square Society, inspiration and mentor to all undergraduate students, Department of Physics and Astronomy, University of Manitoba. You are dearly missed, my friend. Be kind always. \\ 

I thank the valuable and insightful reviews from the anonymous referees, whose critical advice substantially helped to streamline this work. I acknowledge colleagues at the Grain Research Centre for providing many interesting discussions on electrical permittivity and related issues. I also thank Andrew Senchuk for discussions of complex quantities in optics, Kelvin Au for a thorough technical proofreading of this manuscript and G\'{e}za Reilly for proofreading the text. 
\\

% Don't change these lines
%\bsp	% typesetting comment
\label{lastpage}


\section{Bibliography}

\begin{thebibliography}{99}

\bibitem{olegReview} Bisnovatyi-Kogan G. \& Tsupko O., 2017, Universe, 3, 3, p57 

\bibitem{olegShadow} Perlick V. \& Tsupko O. Yu., 2022, Phys. Rep., 947, 1-39

\bibitem{erMao} Er X., Mao S., 2014, MNRAS, 437, 3, 2180-2186

\bibitem{magnetosphere1} Goldreich P., Julian W. H., 1969, ApJ, 157, 869

\bibitem{fallback} Muslimov A., Page D., 1995, ApJ, 440, L77

\bibitem{pageNS} Page D., Beznogov M. V., Garibary I., Lattimer J.M, Prakash M., Janka H.-T., 2020, ApJ, 898, 2, 125

\bibitem{shapiro} Shapiro S. L. \& Teukolsky S. A., ``Black Holes, White Dwarfs, and Neutron Stars: The Physics of Compact Objects'', John Wiley and Sons, 1983. 

\bibitem{synge1} Synge, J. L., 1960, "Relativity: The General Theory", North-Holland, Amsterdam

\bibitem{OG_GR_plasma} Muhleman O. D., Johnston I. D., 1966, Phys. Rev. Lett., 17(8), 455-458 

\bibitem{volkerBook} Perlick V., 2000, ``Ray Optics, Fermat's Principle, and Applications to General Relativity''. Springer-Verlag, Heidelberg, Germany

\bibitem{oleg2} Bisnovatyi-Kogan G. S., Tsupko O. Yu., 2015, Plas. Phys. Rep. 41, 7, 562-581 

\bibitem{olegPlasmaTimeDelay} Bisnovatyi-Kogan G. S., Tsupko O. Yu., 2023, MNRAS, 524, 2, 3060-3067

\bibitem{bozzaOleg} Fabiano F., Bozza V., Tsupko O. Yu., 2024, preprint(arxiv:2406.07703)

% Gravitational weak lensing by a naked singularity in plasma 
\bibitem{exotic1} Atamurotov, F., Ghosh, S. G., 2022, EPJP, 137, 6, 662

% Gravitational weak lensing by black hole in Horndeski gravity in presence of plasma
\bibitem{exotic2} Atamurotov F., Sarikulov F., Abdujabbarov A. \& Ahmedov B., 2022, EPJP, 137, 3, 336

% Light deflection by squashed Kaluza-Klein black holes in a plasma medium
\bibitem{exotic3} Matsuno K., 2021, Phys. Rev. D, 103, 4

\bibitem{olegHillsHoles} Tsupko O. Yu. \& Bisnovatyi-Kogan, G.S., 2020, MNRAS, 491, 4, 5636-3649

\bibitem{dustFreqs} Ling Z., Zhang S. N., 2011, Earth, Planets and Space, 63, 10, 1047-1050

\bibitem{dustEnvironments} Blumer H., Safi-Harb S., 2020, ApJL, 904, 2

\bibitem{J119} Blumer H., Safi-Harb S., McLaughlin M. A., 2017, ApJL, 850:L18 (6pp)

\bibitem{magnetarDust1} Tiengo A. et al., 2010, ApJ 710:227-235

\bibitem{drudeOrig} Drude P., 1900, Ann der Phys. 306, 3, 566-613

\bibitem{eikonal2} Born M. \& Wolf E., 1959, Principles of Optics, Pergamon Press, New York.

\bibitem{rayTracingAxions} McDonald J. L., Witte S. J., 2023, preprint (arxiv:2309.08655)

\bibitem{OG1} Breuer R. A., Ehlers J., 1980, Proc. R. Soc. Lond., A 370, 389-406 

\bibitem{OG2} Breuer, R. A., Ehlers J., 1981, Astron. Astrophys. 96, 293-295

\bibitem{OG3} Noonan T. W., 1982, ApJ, 262, 344-348 

\bibitem{drudeXRay} Smith R. K. \& Dwek E., 1998, ApJ, 503, 831-842

\bibitem{xrayDrude2} Corrales L. R., Garcia J., Wilms J., Baganoff F., 2016, MNRAS, 458, 2, 1345-1351

\bibitem{CP1} Lindquist R. W., 1966, Ann. of Phys., 37, 487-518

\bibitem{CP2} Bicak J., Hadrava P., 1975, Astron \& Astrophys., 44, 389-399

%%%%%%
\bibitem{reissPerlmutter} Reiss A. G., Filippenko A. V., Challis P., et al., 1998, ApJ, 116:1009-1038

\bibitem{chenKantowski1} Chen B. \& Kantowski R., 2008, Phys. Rev. D, 78, 044040

\bibitem{chenThesis} Chen B., 2009, Ph.D. thesis, "Cosmology with a Dark Refraction Index", U. Oklahoma, OK, USA

\bibitem{chenKantowski2} Chen B. \& Kantowski R., 2009, Phys. Rev. D, 79, 104007

\bibitem{dustCosmology} Lima J. A. S., Cunha J. V. \& Zanchin V. T., 2011, ApJL 742:126 (5pp)

\bibitem{chen13} Chen, J., Wu, P.-X., Yu, H.-W., Li, Z.-X., 2013, Res. Astron. Astrophys., 13 635

\bibitem{ck20}  Vavrycuk, V., Kroupa, P., 2020, MNRAS, 497, 1, 378-388

%%%%%%

\bibitem{olegRadioLens} Bisnovatyi-Kogan G. S., Tsupko O. Yu., 2010, MNRAS, 404, 4, 1790-1800

\bibitem{reducedWavelength} Mashhoon B., 1986, Phys. Lett. A., 122, 67

\bibitem{oleg24} Bezdekova, B., Tsupko O., Pfeifer, C., 2024, preprint (arxiv:2403.16842)

\bibitem{mtw} Misner C. W., Thorne K. S. \& Wheeler J. A., 1973, ``Gravitation'', W. H. Freeman, Princeton University Press 

\bibitem{gordon1923} Gordon W., 1923, Ann. Phys. (Leipzig) 72, 421

\bibitem{galloOpticalMetric} Crisnejo G., Gallo E., Villanueva, J. R., 2019, Phys. Rev. D, 100, 4, 044006

\bibitem{spinHall} Andersson L., Oancea M. A., 2023, Class. QUantum. Grav. 40, 154002

\bibitem{carroll} Carroll B. W. \& Ostlie D. A., ``An introduction to Modern Stellar Astrophysics'', Pearson, 2nd Edition, 2006

\bibitem{rybickiLightman04} Rybicki \& Lightman, ``Radiative Processes in Astrophysics, 2004, Wiley-VCH Verlag GmbH 62 Co. KGaA, Weinheim

\bibitem{vincent11} Vincent F. H., Paumard T., Gourgoulhon E., Perrin G., 2011, Class. Quantum Grav., 28, 22, 225011

\bibitem{olegGeneral} Tsupko O. Yu., 2021, Phys. Rev. D., 103, 10, 104019 

\bibitem{dempsey} Dempsey D., Dolan S. R., 2016, Int. J. Mod. Phys. D, 25, 9, 1641004

\bibitem{linKK1} Boukamp, B.A., 1995, J. Electrochem. Soc. 142, 6, 1885-1894

\bibitem{linKK2} Sadkowski A., Dolata M., Diard J. P., 2004, J. Electrochem. Soc., 151, 1, E20

\bibitem{linKK3} Schönleber M., Klotz D. \& Ivers-Tiffée E., 2014, Electrochimica Acta 131, 20-27

\bibitem{kravtsov99} Kravtsov, Yu. A., Forbes G. W., Asatryan A. A., 1999, Theory and Applications of Complex Rays, Progress in Optics XXXIX, Elsevier, Ed. E. Wolf. 

\bibitem{marklund} Marklund M. et al., 2005, Phys. of Plasmas, 12, 7, 4 pp

\bibitem{draine2021} Draine B. T. \& Hensley B. S., 2021, ApJ, 909:94 (21 pp)

\bibitem{oleg_volker} Perlick, V., Tsupko, O. Yu. \& Bisnovatyi-Kogan, G. S., 2015, Phys. Rev. D, 92, 104031

\bibitem{olegRadioSpectrometer} Bisnovatyi-Kogan G. S., Tsupko O. Yu., 2009, Gravitation and Cosmology, 15, 1, 20-27






- 1





\bibitem{narayan} Narayan R., Bartelmann M., 1996, Lectures on Gravitational Lensing, arXiv:astro-ph/9606001

\bibitem{integrals} Gradshtein I.S., Ryzhik I.M., 2007, Tables of Integrals, Series, and Products, 7th Edition, Academic Press, Elsevier, Burlington, MA, USA.

\bibitem{schneider} Schneider P., Ehlers J., Falco E. E., 1992, Gravitational Lenses, XIV, 560 pp. Springer-Verlag Berlin, Heidelberg, New York

\bibitem{rogersEr2018} Rogers A., Er X., 2018, MNRAS, 475, 867-878

\bibitem{linKKBackup3} Schönleber M., \& Ivers-Tiffée E., 2015, Electrochemistry Communications 58, 15-19

\bibitem{linKK_esteban} Esteban J. M. \& Orazem M. E., 1991, J. Electrochem. Soc., 138, 1, 67

\bibitem{linKK_agarwal1} Agarwal P., Orazem M. E. \& Garcia-Rubio L. H., 1992, J. Electrochem. Soc., 139, 1917

\bibitem{heyl1} Shaviv N. J., Heyl J. S. \& Lithwick Y., 1999, MNRAS, 306, 2, 333-347

\bibitem{magBH2} Ahmedov B., Turimov B. Stuchlik Z., Tursunov Z., 2019, Int. J. Mod. Phys. Conf. Series, 49, 1960018

\bibitem{magBH1} Turimov B., Ahmedov B., Abdujabbarov A., Bambi C., 2019, Int. J. Mod. Phys. D, 28, 16, 2040013-187

\bibitem{rogers15} Rogers A., 2015, MNRAS, 451, 1, 17-25

\bibitem{NSPulseOrig} Pechenick K. R., Ftaclas C., Cohen J. M., 1983, ApJ, 274:846-857

\bibitem{NSPulseCharged} Dabrowski M. P., \& Osarczuk J., 1995, aoss 229(1):139-155

\bibitem{NSPulseTurolla} Turolla R. \& Nobili L., 2013, ApJ, 768(2):147

\bibitem{galloNSA} Briozzo G. \& Gallo E., 2023, EPJC, 83, 2, 165

\bibitem{Xray1} Pintore F., Mereghetti S., Tiengo A., et al., 2017, MNRAS, 467, 3, 3467-3474

\bibitem{Xray2} Esposito P., Tiengo A., Rea N., et al., 2013, MNRAS, 429, 4, 11

\bibitem{perlick24} Perlick V., 2023, Astron. Rep., 67, 2, S102-S107

\bibitem{EHT} Mann C. R., Richer H., Heyl J., Anderson J., Kalirai J., Caiazzo I., Mohle S. D., Knee A., and Baumgardt H., Event Horizon Telescope Collaboration, 2019, ApJL, 875, 1

\bibitem{photonMass1} Kulsrud R. \& Loeb A., 1992, Phys. Rev. D., 45, 2

\bibitem{photonMass2} Tsupko O. Yu. \& Bisnovatyi-Kogan G. S., 2013, Phys. Rev. D., 87, 124009

\bibitem{ESE1} Clegg A. W., Fey A. L., Lazio T. J. W., 1998, ApJ, 496, 1, 253-266

\bibitem{ESE2} Fiedler R. L., Dennison B., Johnston K. J., Hewish A., 1987, Nature, 326, 6114, 675-678

\bibitem{ErESE} Wagner J., Er X., 2020, preprint, arXiv:2006.16263

\bibitem{erFRB} Er X., Yang Y.-P. \& Rogers A., 2020, ApJ, 889, 2

\bibitem{erFRB2} Er X., Mao S., 2022, MNRAS 516, 2, 2218-2222

\bibitem{DW86A} Deguchi S., Watson W. D., 1986, Phys. Rev. D., 34, 6, 15, 1708-1718

\bibitem{DW86B} Deguchi S., Watson W. D., 1986, ApJ, 307, 30

\bibitem{takahashiWave1} Takahashi R., Suyama T., Michikoshi S., 2005, A\&A, 438, 1, L5-L8

\bibitem{nakamura_deguchi_1999_review} Nakamura T.T., Deguchi S., 1999, Prog. Theoretical Phys. Suppl. 133, 137-153

\bibitem{takahashi_nakamura_wave_effects} Takahashi R., Nakamura T., 2003, ApJ, 595, 2, 1039-1051 

\bibitem{selmke} Selmke M., Cichos F., 2013, Am. J. Phys. 81, 6, 405-413

\bibitem{evans1} Am. J. Phys, 64, 11, 1404-1415

\bibitem{evansFMA} Evans J. \& Rosenquist M., 1986, Am. J. Phys, 54, 10, 876-883

\end{thebibliography}
\end{document}